\documentclass[10pt,a4paper]{article}

\usepackage{amsmath}
\usepackage{amsfonts}
\usepackage{amssymb}
\usepackage{pdfsync}
\usepackage{epsfig}
\usepackage{graphicx}
\usepackage{subfigure}
\usepackage{cite} 
\usepackage[english]{babel}

\def\nn{\nonumber}
\def\Z{\zeta}
\def\l{l_x}
\def\E{\eta}
\def\Thet{\Omega}
\def\Standard{Standard}
\def\Alternative{Alternative}
\def\OO{\mathcal{O}}

\begin{document}

\newcommand{\be}{\begin{equation}}\newcommand{\ee}{\end{equation}}
\newcommand{\bea}{\begin{eqnarray}} \newcommand{\eea}{\end{eqnarray}}
\newcommand{\ba}[1]{\begin{array}{#1}} \newcommand{\ea}{\end{array}}
\renewcommand{\thesection}{\arabic{section}}
\renewcommand{\theequation}{\thesection.\arabic{equation}}

\renewcommand{\thesubfigure}{\thefigure.\arabic{subfigure}} 
\makeatletter 
\renewcommand{\@thesubfigure}{\text{Figure\ }\thesubfigure} 
\renewcommand{\p@subfigure}{} 
\makeatother

\thispagestyle{empty}

\renewcommand{\thefootnote}{\fnsymbol{footnote}}
\setcounter{page}{1}
\setcounter{footnote}{0}
\setcounter{figure}{0}

\def\nn{\nonumber}
\def\Z{\zeta}
\def\l{l_x}
\def\E{\eta}
\def\Thet{G}
\def\Standard{Standard}
\def\Alternative{Alternative}
\def\OO{\mathcal{O}}

\bigskip

\rightline{June, 2012}

\vspace{1cm}

\begin{center}

{\LARGE\bf  
Holographic Superconductors in a Cohesive Phase}\\
\bigskip
{\Large\bf The Retrograde Condensate paid back}\\
\vspace{2cm}

{\large\sc Francesco Aprile\footnote{\normalsize{Email address: dascancellare@gmail.com or aprile@ecm.ub.es}} }\\
\bigskip
{\it  Institute of Cosmos Sciences and Estructura i Constituents de la Materia\\
Facultat de F{\'\i}sica, Universitat de Barcelona, Av. Diagonal 647,  08028 Barcelona, Spain}\\
\vspace{0.25cm}
{\it and}\\
\vspace{0.25cm}
{\it Perimeter Institute for Theoretical Physics, Waterloo, Ontario N2L 2Y5, Canada}\\
\bigskip
\vspace{2cm}

\end{center}

\begin{abstract}
\smallskip
\noindent
{\normalsize
We consider a four-dimensional $\mathcal{N}=2$ gauged supergravity coupled to matter fields. The model is obtained by a $U(1)$ gauging of a charged hypermultiplet and therefore it is suitable for the study of holographic superconductivity. The potential has a topologically flat direction and the parameter running on this ``moduli space'' labels the new superconducting black holes. The phase diagram of the theory is studied and the zero temperature solutions are constructed. The model has rich dynamics. The retrograde condensate is just a special case in the new class of black holes. The calculation of the entanglement entropy makes manifest the properties of a generic solution and the superconductor at zero temperature is in a confined cohesive phase. The parameter running on the topologically flat direction is a marginal coupling in the dual field theory. We prove this statement by considering the way double trace deformations are treated in the $AdS/CFT$ correspondence. Finally, we comment on a possible connection, in the context of gauge/gravity dualities, between the geometry of the scalar manifold in $\mathcal{N}=2$ supergravity models and the space of marginal deformations of the dual field theory.} 
\end{abstract}

\clearpage

\tableofcontents


\section{Introduction}

Over the past few years the $AdS/CFT$ correspondence has been one of the most studied subjects in theoretical physics \cite{Maldacena:1997re}. The duality has lead to the foundation of the holographic principle which relates a certain class of field theories at strong coupling to general relativity in asymptotically $AdS$ spaces. The precise prescription for computing physical observables of the field theory through classical gravity makes the duality a powerful tool to deal with strong coupling dynamics \cite{Witten:1998qj,Gubser:1998bc}. More generally, holographic techniques have been successfully applied on a vast number of problems, ranging from the quark-gluon-plasma to the most recent and challenging aspects of condensed matter physics \cite{Hartnoll:2009sz, McGreevy:2009xe}. 

The call for holography as guiding principle in condensed matter systems, comes from the lack of a field theory explanation for the properties of several many body systems. The physics underlying high-$T_c$ superconductors and strange metals seems to evade the well-established result that the Fermi liquid theory of fermions governs the low energy excitations of a Fermi surface, independently of the strength of the electron interaction at the lattice scale \cite{Wilson:1993dy,Polchinski:1992ed}. On the other hand, this result remains an outcome of RG flow techniques in a regime in which perturbative quantum field theory can be applied. Holographic theories have been introduced as an alternative framework to overcome this issue. They naturally account for strong coupling physics and quite surprisingly, a recent work has shown that in a ``realistic''  holographic setting, some properties of the cuprates come for free. Thus, high-$T_c$ superconductors and strange metals may belong to universality classes that admit an holographic description \cite{Hartnoll:2011fn,Faulkner:2011tm,Jensen:2011su,Faulkner:2010da,Horowitz:2012ky}.

The model presented in this paper belongs to the class of theories introduced as holographic superconductors \cite{Hartnoll:2008kx,Horowitz:2010gk, Herzog:2009xv}. The strongly coupled field theory is studied through its holographic dual and the field content of the Lagrangian, in the simplest case, consists of gravity, a complex scalar and a gauge field. Black hole geometries which support electric flux are dual to states of the field theory with finite density of matter. These geometries are generically unstable at low temperature and spontaneously develop a scalar hair which is responsible for a field theory operator to condense. Despite the simplicity of the model, the condensation seems to be a universal feature of charged gravitational backgrounds whenever a charged scalar field can be turned on. The only wrinkle in this story is given by the existence of the retrograde condensate, first pointed out in \cite{GubStrings}. This is an exotic and not thermodynamically favored phase of the theory, which has the odd characteristic of showing a condensate only for temperatures above a certain critical temperature. Nevertheless, it has been found in models coming from string theory. Motivated by this last observation, we want to point out that the retrograde condensate has been discovered in models obtained by truncating a complicated supergravity theory to a simpler sector. Then, one might suspect that the simplicity of those truncations is hiding part of the story. In other words, we chase the idea that interesting physics comes together with the retrograde condensate.

Our model is an $\mathcal{N}=2$ supergravity theory coupled to a charged hypermultiplet and it can be embedded in the four dimensional $\mathcal{N}=8$ supergravity. It includes two special solutions already discussed in the literature: a well-know holographic superconductor and a retrograde condensate. The main point of this paper is the construction of a class of hairy black holes whose condensates interpolate between the above mentioned special solutions. The new class of superconducting solutions present remarkable and novel dynamical aspects. In this sense, the idea that interesting physics was related (even if not directly) to the retrograde condensate is paid back.

$\mathcal{N}=2$ supergravity coupled to hypermultiplets is quite special. Hyperscalars are associated to coordinates on a quaternionic manifold and supersymmetry fixes completely the form of the Lagrangian. In this paper we analyze the dynamics of the two charged hyperscalars which parametrize the coset space $SU(2,1)/U(2)$. The scalar potential is non trivial: it has a local maximum at the origin, a saddle point and two tachyonic directions, one related to the saddle point and the other one related to the retrograde condensate; they both join at the boundary of the scalar manifold. 
The coset space is a two dimensional complex manifold with quaternionic structure but can also be described as an open ball with a non trivial embedding. Because of this topological property, a flat direction shows up in the scalar potential. An angle labels the set of $AdS$ vacua associated with the flat direction and we call this parameter $\theta_{\infty}$. Then, the new class of hairy black holes is constructed by demanding that the superconducting solutions approach the asymptotically AdS vacuum at a fixed value of the parameter $\theta_{\infty}$. Tuning $\theta_{\infty}$ along the flat direction changes the shape of the condensate. The well-know example of superconductor is found for $\theta=0$ whereas increasing the parameter up to $\theta=\pi/2$ turns the condensate into a retrograde condensate. In between, a cross-over of first order phase transitions appears. The phenomenology of these condensates is clarified by Figure \ref{Figura1A} and we refer to them as ``interpolating solutions''. The features of the condensates are not only qualitative like in a phenomenological theory: we quantify the properties of the interpolating solutions cooling the system down to zero temperature.

In order to gain some intuition about the informations that the holographic principle provides when the theory is considered at zero temperature, we would like to stress the following important point. According to the Wilsonian renormalization group approach, low energy physics of a theory is obtained by integrating out the microscopic degrees of freedom down to the energy scale of interest. This idea is geometrically realized in theories with a gravity dual where the bulk radial coordinate represents the resolution scale of the dual field theory. In this case the short distances degrees of freedom of the field theory are associated to near boundary regions of the gravitational background whereas the low energy physics emerges from the shape of the geometry at small radial scales \cite{Faulkner:2010jy}. Holographic superconductors at zero temperature can be described in terms of RG flow dynamics. In the AdS/CFT correspondence we are implicitly assuming that the dual field theory lives on the conformal boundary of an asymptotically $AdS$ space. In domain wall geometries between two $AdS$ spaces, conformal invariance is also emergent at low energies \cite{Gubser:2009gp}. In our extremal interpolating solutions a new scale appears in the gravity background and the geometry in the far IR collapses. We call this new scale $C_{\eta}$. More precisely, below $C_{\eta}$ the bulk background has the shape of a cone all the way down to the tip. This characteristic brings to mind the gravity description of a confining theory as a cigar geometry \cite{Witten:1998zw}. Then, in the spirit of an effective field theory, we may wonder what kind of theory describes the low energy properties of the superconducting ground state dual to an interpolating solution. The holographic geometry suggests that this IR effective theory is a confining theory. To confirm this picture we study the behavior of the entanglement entropy.

We recall that the entanglement entropy is a non local quantity which measures the amount of entanglement among the degrees of freedom of a ground state. It has a natural formulation in quantum many-body systems such as quantum spin chains \cite{Latorre:2003kg, Calabrese:2005zw}. The system is characterized by a density matrix $\mathbb{X}$ and the entanglement entropy is given by the von Neumann entropy associated to the reduced density matrix on a certain space domain $\mathcal{E}$,
\be
S=-\mathrm{Tr}\ \mathbb{X}_\mathcal{E}\log\mathbb{X}_\mathcal{E}\ ,\qquad \mathbb{X}_{\mathcal{E}}=\mathrm{Tr}_{\mathcal{E}^c}\ \mathbb{X}\ .
\ee
The holographic dual of this calculation is understood as the analogous of the Bekenstein-Hawking entropy for a bulk surface which has $\partial\mathcal{E}$ as its boundary \cite{Nishioka:2009un}. The key point is that a transition similar to a confinement/deconfinement transition (see for example \cite{Mateos:2007ay}) takes place when the topology of the bulk surface changes from connected to disconnected. The result of our calculation shows that the new scale appearing in the zero temperature interpolating solution, sets the characteristic length $\ell$ of the confinement/deconfinement transition. We use this transition as a measure of cohesion of the superconducting ground state. Indeed, in our zero temperature superconducting ground state, we expect that correlation lengths are finite and proportional to $C_{\E}$. This intuition agrees with the idea that the superconductor is better described as a cohesive state in a strong coupling regime \cite{Hartnoll:2012ux}.  Is this an indication  that the superconductor will remain in a ``glue'' state at all temperatures? We believe that the answer is given by the order of the phase transition. A first order phase transition suggests that correlation lengths are finite also at the critical temperature. 
Our model is the first example of a field theory in which a tunable parameter, the angle $\theta_{\infty}$, accounts for all these phenomena. 

The confinement/deconfinement transition is the natural consequence of the geometry capping off below the new scale $C_{\E}$. We want to emphasize that the shape of the background in the far IR is the response (or the back-reaction) of the geometry to the presence of a UV mixing between the two charged scalars of the hypermultiplet. In the supergravity description this mixing is due to the non trivial $\theta_{\infty}$. Interestingly, a similar mixing also relates the condensates which are dual to the two hyperscalars. This aspect is a remarkable outcome of the $AdS/CFT$ scheme. Then, the field theory interpretation of the boundary data of our interpolating solutions gets interesting. Indeed, it is possible to prove that the parameter $\theta_{\infty}$ is the marginal coupling of a double trace deformation in the dual field theory. However, the deformation does not break conformal invariance and thus another mechanism must be responsible for the RG flow towards the IR confining phase. 
By considering the supergravity potential, it is evident that, in the classical picture, the mixing angle $\theta_{\infty}$ drives the hyperscalars along the tachyonic directions and therefore, towards the new IR geometry. We identify the field theory counterpart of this argument in the mixing between the two condensates: conformal invariance in the UV is broken and a relevant deformation initiates the RG flow. In particular, the same scale $C_{\eta}$ which measures the conical shape of the background, also measures how far an interpolating solution is from the IR conformal fixed point dual to the domain wall geometry.

Several features play an important role in understanding the dynamics of the interpolating condensates. They are all tied up: the shape of the new extremal geometry, the confinement/deconfinement transition and the holographic RG flow interpretation. We believe that every aspect of our model has its roots in the properties of the scalar manifold and therefore, in the $\mathcal{N}=2$ theory under consideration. Ultimately, in its string theory origin. Unfolding these properties in the context of holographic superconductivity increase the expectations that the $AdS/CFT$ correspondence represents a fruitful approach to condensed matter physics.

\smallskip

The organization of the paper is as follows. In section \ref{S1} we describe the main ingredients of the model and we define the holographic superconductor ansatz. In section \ref{S2} we discuss the details of the supergravity potential explaining the features of the topologically flat direction. We construct a more general class of hairy black holes presenting the numerical results for the finite temperature and the extremal solutions. We calculate the entanglement entropy to make manifest the physical properties of these new solutions. In section \ref{S3} we consider the field theory interpretation of our interpolating solutions. Using the holographic renormalization approach we explain how the parameter running on the topologically flat direction is mapped to a marginal coupling. Then, we give a detailed analysis of the RG flow of the theory. Finally, we relate the geometrical feature of the supergravity model to the existence of the marginal deformation in the dual field theory.

\section{The model \label{S1}}
We consider the bosonic sector of $4D$ $\mathcal{N}=2$ supergravity theory coupled to matter fields for a special value of the gauging parameter. The procedure to obtain the Lagrangian is illustrated in \cite{Andrianopoli:1996vr,Ceresole:2001wi}. The general structure of the theory includes: $n_V$ vector multiplets, $n_H$ hypermultiplets and the gravity multiplet. The scalars parametrize a manifold which is the direct product of a special K\"ahler manifold $\mathcal{SK}(n_V)$ and a Quaternionic manifold $\mathcal{Q}(n_H)$. We analyze a simple model in this class of theories retaining only the gravity multiplet and one hypermultiplet. The quaternionic manifold is $SU(2,1)/U(2)$ and the gauge symmetry is obtained by gauging a $U(1)$ killing vector in the isotropy group $SU(2)\times U(1)$. The model is completely specified by the geometrical construction however, the gauging is not unique and the more general Lagrangian depends on the choice of a single gauging parameter. This parameter picks a $U(1)$ direction in the isotropy group $SU(2)\times U(1)$. We consider the model obtained by gauging the $\sigma_3$ direction in the $SU(2)$ subgroup. This particular model can be embedded in a consistent truncation of the $\mathcal{N}=8$ supergravity \cite{Bobev:2011rv}.

The action of our model is given by, 
\be\label{LWarner}
\mathcal{L}=\frac{1}{2\kappa^2}\Big(\ \mathcal{R}-\frac{1}{2}F_{\mu\nu}F^{\mu\nu}\ -\ 2\Thet^{MN}(\mathcal{D}^{\mu}\Z)_M(\mathcal{D}_{\mu}\overline{\Z})_N-\mathcal{P}\ \Big)\ ,
\ee\\
where $F=dA$ is the field strength of the $U(1)$ gauge field $A$.
The quaternionic manifold $SU(2,1)/U(2)$ is a $2$-dimensional complex manifold endowed with a metric $\Thet^{MN}$ originating from a  K\"ahler potential $\mathcal{K}$,
\be
\mathcal{K}=-\frac{1}{2}\ln(1-|\Z_1|^2-|\Z_2|^2)\ .
\ee
In terms of the complex scalars $\Z_1$ and $\Z_2$ the metric is explicitly written in the form,\\
\bea 
\nn
d\Thet^2 & = & \Thet^{MN}d\Z_M\overline{d\Z}_N\ ,\\
\nn\\
d\Thet^2 & = & 
\frac{d\Z_1\overline{d\Z}_1+d\Z_2\overline{d\Z}_2}{(1-|\Z_1|^2-|\Z_2|^2)}
+\frac{(\overline{\Z}_1d\Z_1+\overline{\Z}_2d\Z_2)(\Z_1\overline{d\Z}_1+\Z_2\overline{d\Z}_2)}{(1-|\Z_1|^2-|\Z_2|^2)^2}\ .\label{metricZ}
\eea
The covariant derivatives on the complex scalars are
\be\label{CovariantDerivative}
D_{\mu}\zeta_1=\partial_{\mu}\zeta_1+\ i g A_{\mu}\zeta_1\ ,\qquad D_{\mu}\zeta_2=\partial_{\mu}\zeta_2-\ i g A_{\mu}\zeta_2\ ,
\ee
and the potential is 
\be
\mathcal{P}=-g^2\ \frac{12-16(|\zeta_1|^2+|\zeta_2|^2)+3(|\zeta_1|^4+|\zeta_2|^4)+10|\zeta_1|^2|\zeta_2|^2}{(1-|\zeta_1|^2-|\zeta_2|^2)}\ .
\ee
As a consequence of the particular gauging of the rigid theory the scalars have opposite charges $\pm g$. It's convenient to write the equations for $\Z_1$ and $\Z_2$ in the general form, 
\bea
\label{zetaC1}
(\nabla_{\mu}-ig A_{\mu})(\nabla^{\mu} -i g A^{\mu})\zeta_1+(\partial_{\mu}\zeta_1-i g A_{\mu}\zeta_1)\mathcal{X}^{\mu}-\ \mathcal{DV}_1\zeta_1=0\ ,\\ 
\nonumber \\ 
\label{zetaC2}
(\nabla_{\mu}-ig A_{\mu})(\nabla^{\mu} -i g A^{\mu})\zeta_2+(\partial_{\mu}\zeta_2-i g A_{\mu}\zeta_2)\mathcal{X}^{\mu}-\ \mathcal{DV}_2\zeta_2=0\ ,
\eea
where we have defined the quantities,
\bea
\mathcal{X}_{\mu}&=&\frac{2}{1-|\zeta_1|^2-|\zeta_2|^2}\Big(\ \bar{\zeta}_1 D_{\mu}\zeta_1+\bar{\zeta}_2 D_{\mu}\zeta_2\ \Big)\ ,\\ 
\nonumber\\ 
\label{DV1}
\mathcal{DV}_1&=&-g^2\ \frac{4-5|\zeta_1|^2-3|\zeta_2|^2}{1-|\zeta_1|^2-|\zeta_2|^2}\ ,\\
\nonumber\\
\label{DV2}
\mathcal{DV}_2&=&-g^2\ \frac{4-5|\zeta_2|^2-3|\zeta_1|^2}{1-|\zeta_1|^2-|\zeta_2|^2}\ .
\eea
The masses of the scalar fields can be read from the expressions of $\mathcal{DV}_1$ and $\mathcal{DV}_2$. They are equal and are given by $m^{2}=-4g^2$.

\subsection{Superconducting black holes: Ansatz}

The $AdS/CFT$ correspondence in the present context is used as a tool to engineer states in the dual field theory with finite density of matter. We briefly recall how the dictionary between bulk fields and boundary operators works in our specific case. On the field theory side a finite density of matter is introduced by giving an expectation value $\langle J^t\rangle$ to the charge density operator for a global $U(1)$ symmetry. The current $J^{\mu}$ associated to the boundary global symmetry is mapped to a bulk Maxwell gauge field $A_{\mu}$ and the charge density operator is dual to the Maxwell gauge potential. The field theory is put at finite temperature by considering black holes geometries. The Hawking temperature of the black hole corresponds to the temperature of the field theory.

The ansatz for the black hole is specified as follows. Requiring invariance under space-rotations and time-translations, the metric and the gauge field can be taken to be, 
\be\label{Ansatz}
ds^2=-f(r)dt^2+\frac{r^2}{L^2}(dx^2+dy^2+dz^2)+\frac{dr^2}{f(r)h(r)^2}\ ,\qquad A=\Phi(r)dt\ .
\ee
Solutions of interest must be asymptotically $AdS$. As usual, the dual field theory lives in the UV region where $r\rightarrow\infty$. In our model the $AdS$ solution corresponds to the case in which all matter fields vanish. The corresponding metric components are $f(r)=r^2/L^2$ and $h(r)=1$, where we have chosen the normalization $g^2L^2=1/2$. 

The holographic dual of a state with finite charge density is a black hole with electric flux at infinity. In this configuration $\Phi(r)$ has a non trivial profile. One candidate to describe this state is the Reissner-Nordstr\"om black hole,
\be
f(r)=\frac{r^2}{L^2}-\frac{M}{r}+\frac{\rho}{4r^2}\ ,\qquad  h(r)=1\ ,\qquad \Phi(r)=\mu-\frac{\rho}{r}\ .
\ee 
It represents an uncondensed phase because the charged scalar fields play no role in this background. The energy of the configuration is proportional to the mass $M$ of the black hole whereas the charge density of the dual field theory is set by value of $\rho$. Looking for hairy black holes we assume the scalar fields to be functions only of the radial coordinate,
\be\label{AltraDefScalar}
\Z_1\ \rightarrow\ z_1(r)\ e^{i\phi_1(r)},\qquad\ \Z_2\ \rightarrow\ z_2(r)\ e^{i\phi_2(r)}\ .
\ee
In this background the equations (\ref{zetaC1}) and (\ref{zetaC2}) are real and therefore the phases $\phi_1$ and $\phi_2$ are constants. We set them to zero for simplicity. Linearizing the equations of motion around the $AdS$ vacuum we find the generic asymptotics, 
\bea
f(r)&=&\frac{r^2}{L^2}-\frac{M}{r}+\ldots\ ,\label{expansF}\\
\nonumber\\
h(r)&=& 1+\ldots\ ,\label{expansH}\\
\nonumber\\
\Phi(r)&=&\mu-\frac{\rho}{r}+\ldots\ ,\label{expansPhi}\\
\nonumber\\
z_i(r)&=&\frac{\mathcal{O}_i^{(1)} }{r}+\frac{\mathcal{O}_i^{(2)} }{r^2}+\ldots\ . \label{expansZeta}
\eea
The fall-off of the scalar fields is determined by the condition
\be\label{delta}
\Delta(\Delta-3)=m^2L^2\ .
\ee
The mass of the scalars is $m^2L^2=-2$ and therefore the solutions are $\Delta=1,2$. In the general case, the boundary behavior for the scalar field can be written in the following form,
\be
\zeta=\frac{\mathcal{O}_i^{(1)} }{r^{3-\Delta}}+\frac{\mathcal{O}_i^{(2)} }{r^\Delta}\ , 
\ee 
where $\Delta$ is the larger of the two solutions of (\ref{delta}). For masses in the range $m^2L^2 > m^2_{BF}+1$, the only normalizable mode is associated to the $\Delta$. In the field theory language the scalar field is mapped to an operator of dimension $\Delta$ and the coefficient $\mathcal{O}^{(2)}$ is the expectation value taken on the state dual to the gravitational background. The coefficient $\mathcal{O}^{(1)}$ (of dimension $3-\Delta $) is interpreted as a source for the dual operator and if the source is non vanishing then the field theory action is modified by a term proportional to
\be\label{modificatio}
\delta\mathcal{S}\propto\int d^3 x\ \mathcal{O}^{(1)}\mathcal{O}^{(2)}\ .
\ee
This scheme goes under the name of ``Standard Quantization''. For masses in the range $m^2_{BF}< m^2L^2 < m^2_{BF}+1$, a slight modification of the action makes both modes normalizable and the role of source and condensate can be reversed \cite{Klebanov:1999tb}. In this case, the field theory operator has dimension $3-\Delta$, the coefficient $\mathcal{O}^{(1)}$ is interpreted as the condensate and $\mathcal{O}^{(2)}$ is associated to the source. If the source is non vanishing then the field theory action is modified by a term similar to (\ref{modificatio}). This reversed interpretation goes under the name of ``Alternative Quantization''. The mass $m^2L^2=-2$ belongs to the range in which the two types of quantizations are possible.

The $U(1)$ global symmetry of the field theory is spontaneoulsy broken by the condensate if the asymptotics of the scalar fields set the sources to zero \cite{Gubser:2008px}. Imposing regularity at the horizon allows a two-parameter family of solutions to the equations of motion. Once we require the spontaneous symmetry breaking condition, the superconducting black hole is specified only by the temperature. We leave further details to the Appendix. It is worthwhile to mention that we always work in the canonical ensemble where the charge density of the system is fixed.
%
%
\subsection{The old and the new}

The Lagrangian (\ref{LWarner}) contains two superconducting black hole solutions already studied in the literature. We briefly review how they are embedded in our model (\ref{LWarner}) and what are their main features.\\

$\bullet\ $ Model I,
\be\label{MWarner1}
\Z_1=\tanh \frac{\E}{2}\ , \qquad \zeta_2=0\ ,\qquad \mathcal{P}=g^2\ \left(\sinh^4\left(\frac{\E}{2}\right)-4\ (2+\cosh \E)\right)\ ,
\ee
corresponds to \cite{Bobev:2011rv}
\be\label{MWarner2}
\mathcal{L}=R-\frac{1}{2}F^2-\frac{1}{2}(\partial\E)^2-\frac{g^2}{2}\sinh^2\E\ A^2-\mathcal{P}\ .
\ee
Choosing the alternative quantization scheme for $\zeta_1$ a superconducting black hole is found for $T_c\approx 0.121$, the phase transition is second order and the solution can be cooled down to zero temperature. It is the thermodynamically preferred phase. At $T=0$ the bulk geometry is nicely understood as a domain wall between two $AdS_4$ regions with different radius. We will say more about this kind of geometry in the following sections.\\  

$\bullet\ $ Model II,
\be\label{MGauntlett1}
\Z_1=\frac{1}{\sqrt{2}}\tanh \frac{\E}{2}\ , \qquad \Z_2=\frac{1}{\sqrt{2}}\tanh \frac{\E}{2}\ , \qquad \mathcal{P}=-4g^2(2+\cosh \E)\ ,
\ee
corresponds to \cite{Donos:2011ut}
\be\label{MGauntlett2}
\mathcal{L}=R-\frac{1}{2}F^2-\frac{1}{2}(\partial\E)^2-2 g^2\ \mathrm{Sinh}^2\left(\frac{\E}{2}\right) A^2-\mathcal{P}\ .
\ee
The critical temperature of the hairy black hole solutions is the same of Model I, $T_c\approx 0.121$, but the superconducting condensate exists only for temperatures above $T_c$. The solution is not thermodynamical preferred and the system always stays in the uncondensed phase. The shape of the curve suggests that the phase should be classify as a ``retrograde condensate''\footnote{Related exotic aspects and a more accurate analysis of GL instabilities associated with a retrograde condensate can be found in \cite{Buchel:2010wk}.}.\\ 

Model I and Model II corresponds to two different scalar configurations however, they share a common feature, $|\Z_1|^2+|\Z_2|^2=\tanh^2(\eta/2)$. This observation suggests that there is a larger class of black holes in our theory\footnote{We thank Jorge Russo for this very important comment.}. In this new class of solutions Model I and Model II are only isolated points. In the next section we present a more sensible parametrization of the scalar manifold that nicely fits with our purpose. The situation is fascinating at least for one reason: it may clarify what kind of deformation relates Model I to the retrograde condensate.

\section{Interpolating solutions: Geometrical description \label{S2}}

The coset space $SU(2,1)/U(2)$ is topologically a ball and can be parametrized by a set of four real fields. 
Consider the following change of variables,
\be\label{interPAnsatz}
\Z_1= \tau \cos\frac{\theta}{2}e^{i(\varphi+\psi)/2}\ , \qquad\qquad \Z_2= \tau \sin\frac{\theta}{2}e^{-i(\varphi-\psi)/2}\ .
\ee
It takes the metric (\ref{metricZ}) into the form,
\be
d\Thet^2=\frac{d\tau^2}{(1-\tau^2)^2}+\frac{\tau^2}{4(1-\tau^2)}(\sigma_1^2+\sigma_2^2)+\frac{\tau^2}{4(1-\tau^2)^2}\sigma_3^2\ ,
\ee
where $\hat{\sigma}_i$ are the standard left invariant one-forms \cite{BrittoPacumio:1999sn},
\bea
\hat{\sigma}_1&=&\cos\psi\ d\theta+\sin\psi\sin\theta\ d\varphi\ ,\nn\\
\hat{\sigma}_2&=&-\sin\psi\ d\theta+\cos\psi\sin\theta\ d\varphi\ ,\nn\\
\hat{\sigma}_3&=&d\psi+\cos\theta\ d\varphi\ .\nn
\eea
The definition of the new variables is restricted to, $r\in [0,1),\ \theta\in [0,\pi),\ \varphi\in[0,2\pi)$ and $\psi\in[0,4\pi)$. We recall that the phases are always irrelevant; they do not appear in the potential and they are constants in the ansatz (\ref{Ansatz}). Defining $\tau=\tanh(\eta/2)$ and $\mathcal{A}=A/\sqrt{2}$, we obtain the effective Lagrangian of the model (\ref{LWarner}),
\be\label{LagDef}
\mathcal{L}=\frac{1}{2\kappa^2}\Big(
\mathcal{R}-
\frac{1}{4}\mathcal{F}^2-
\frac{1}{2}\sinh^2\left(\frac{\eta}{2}\right)(\partial \theta)^2-\frac{1}{2}(\partial\eta)^2-
\frac{1}{2}J(\eta,\theta)\mathcal{A}_{\mu}\mathcal{A}^{\mu}-\ \mathcal{P}(\eta,\theta)\ \Big)\ .
\ee
The potential and the coupling $J(\eta,\theta)$ in the field variables $\{\eta,\theta\}$ are,\\
\bea
\mathcal{P}(\eta,\theta)&=&g^2 \left(\  \sinh^4\left(\frac{\E}{2}\right)\ \cos^2\theta\ - 4\ (2+\cosh\eta)\right)\ ,\\
\nn\\
J(\eta,\theta)&=&2g^2\sinh^2\left(\frac{\E}{2}\right)\ \left(1+\cos^2\theta\ \sinh^2\left(\frac{\E}{2}\right)\right)\ .
\eea
The equations of motion are 
\begin{flalign}
	\label{Eqchi}
\frac{h'}{rh}+\frac{1}{4}\E'^2+\frac{1}{4}\sinh^2\left(\frac{\E}{2}\right)\theta'^2+J(\E,\theta)\frac{\Phi^2}{4 f^2 h^2}=0&\phantom{1}&\\\nn\\
	\label{Eqg}
-\frac{1}{4}\E'^2-\frac{1}{4}\sinh^2\left(\frac{\E}{2}\right)\theta'^2+\frac{\Phi'^2}{4 f}+\frac{f'}{rf}+
\frac{1}{r^2}+\frac{1}{2 f h^2}\mathcal{P}(\E,\theta)-J(\psi,\varphi)\frac{\Phi^2}{4 f^2 h^2}=0&\phantom{1}&\\\nn
\\
	\label{EqPhi}
\Phi''+\Big(\frac{2}{r}+\frac{h'}{h}\Big)\Phi'- J(\E,\theta)\frac{\Phi}{f h^2}=0&\phantom{1}&\\\nn\\
	\label{Eqpsi}
\E''+\left(\frac{2}{r}+\frac{f'}{f}+\frac{h'}{h}\right)\E'-\frac{1}{4}\sinh\E\ \theta'^2+\partial_{\E}J(\E,\theta) \frac{\Phi^2}{2f^2 h^2}-
\frac{1}{f h^2}\partial_{\E}\mathcal{P}(\E,\theta)=0&\phantom{1}&\\\nn\\
	\label{Eqtheta1}
\theta''+\left(\frac{2}{r}+\frac{f'}{f}+\frac{h'}{h}\right)\theta'+\coth\left(\frac{\E}{2}\right)\E'\theta'+\frac{1}{\sinh^2\left(\E/2\right)}
\left(\partial_{\theta}J(\E,\theta) \frac{\Phi^2}{2 f^2 h^2}-\frac{1}{f h^2}\partial_{\theta}\mathcal{P}(\E,\theta)\right)=0&\phantom{1}&\\\nn
\end{flalign}

We discuss what values of $(\eta,\theta)$ are critical points of the potential $\mathcal{P}(\E,\theta)$. We need to solve the two algebraic equations, 
\bea
\partial_{\E}\mathcal{P}=-g^2 \sinh\E\ \Big(4-\cos^2\theta\ \sinh^2\left(\frac{\E}{2}\right)\Big)=0\label{cond1}\\
\nn\\ 
\partial_{\theta} \mathcal{P}=-2 g^2 \sin2\theta\ \sinh^4\left(\frac{\E}{2}\right)=0\label{cond2}
\eea
A simple solution is found for $\E=0$. For this case $\partial_{\E}\mathcal{P}$ and $\partial_{\theta} \mathcal{P}$ vanish automatically no matter what value of $\theta\in [0,\pi)$ we fix. Since the functions $\mathcal{P}(\E,\theta)$ and $J(\E,\theta)$ are $\pi/2$-periodic it's sufficient to consider the following set,
\be
\mathcal{M}_{\theta}=\{(\eta=0, \theta)\ |\ \theta \in [0,\pi/2]\}\ .
\ee
The set $\mathcal{M}_{\theta}$ resembles a flat direction but not in the usual sense\footnote{In supersymmetric theories, as well as in supergravity theories, a classical flat direction in the potential is parametrized by a massless field whose expectation value sets the mass spectrum of the theory.}. Technically, the reason is that both scalars, $\Z_1$ and $\Z_2$, have non zero mass, $m^2L^2=-2$. What happens is that $\mathcal{M}_{\theta}$ appears as a flat direction only in the variables $\{\eta,\theta\}$. 
Under our change of coordinates (\ref{interPAnsatz}), $\eta=0$ (or $\tau=0$) implies $\Z_1=\Z_2=0$ and the whole $\mathcal{M}_{\theta}$-set is mapped to the origin $\Z_1=\Z_2=0$. Therefore, the interpretation of $\theta$ is obtained by considering the ratio $\Z_2/\Z_1$ which gives a term proportional to $\tan(\theta/2)$. The choice of $\theta$ fixes the direction of departure from the origin towards the boundary. This is a consequence of the topology of $SU(2,1)/U(2)$ that can be seen as the open ball in $\mathbb{C}^2$. Indeed, the coordinates (\ref{interPAnsatz}), for fixed $\tau\neq 0$, represent the Hopf fibration of the three-sphere. We refer to $\mathcal{M}_{\theta}$ as a topologically flat direction. 

\begin{figure}[t]
  \centering
  \includegraphics[scale=.7]{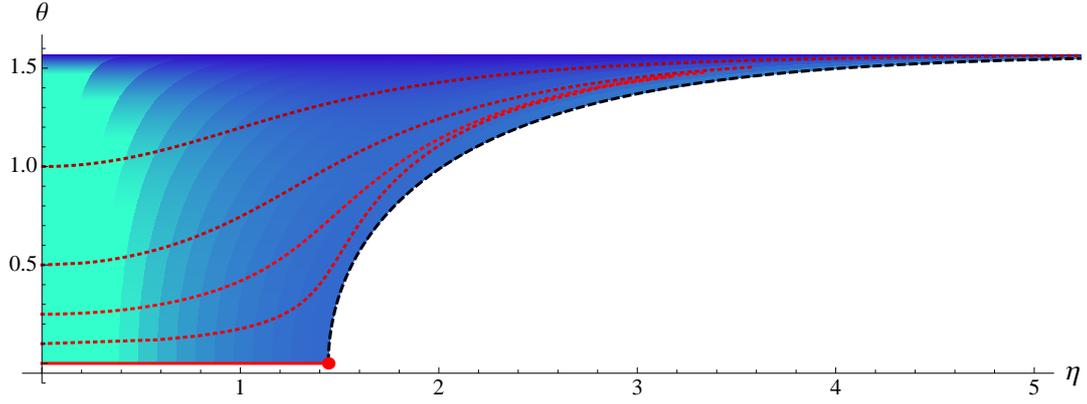}
  \caption{\it \small{Density plot of the potential $\mathcal{P}$. The potential gets steeper as the color becomes darker. The red dot is the saddle point $\theta=0$, $\eta=2\ {\rm arccosh} \sqrt{5}$. For convenience the horizontal axis has been rescaled by a factor $0.5$. The dashed black line is the set $\{ (\theta,\eta)\ |\  \partial_\eta\mathcal{P}=0\}$. We anticipate the results of the section \ref{S21}. The red lines, both dotted and solid, are the horizon values of the functions $\theta(r)$ and $\eta(r)$ when the superconducting black holes are constructed. In particular the value of $\theta_{\infty}$, which specifies the solution, can be read on the vertical axis: it coincides with $\theta(r_h)$ when $\E(r_h)=0$. On each line, the boundary conditions which define a superconducting black hole are satisfied. The figure shows that for the case $\theta(r_h)\neq 0$ the superconductor is driven towards $\theta(r_h)=\pi/2$.
}}
	\label{Figura0}
\vspace{-.25cm}
\end{figure}

It is useful to describe the behavior of the potential for a generic value of $\theta$. In Figure \ref{Figura0} we show a density plot of $\mathcal{P}(\E,\theta)$. The black dashed line is the set of points $\{ (\theta,\eta)\ |\  \partial_\eta\mathcal{P}=0\}$. This line separates the plane $(\E,\theta)$ in two regions according to the sign of $\partial_\eta\mathcal{P}$. In the colored region this derivative is negative and the potential decreases. In the white region the potential increases and it behaves like $\mathcal{P}\approx\exp 2\E$ in the large $\E$ limit. In summary, Figure \ref{Figura0} shows that along the slices of constant $\theta\neq\pi/2$ the potential is bounded from below but, on the slice $\theta=\pi/2$, the potential is negative definite and decreases like $\mathcal{P}\approx-\exp \E$. A second isolated critical point $\mathfrak{S}$ exists for $\theta=0$ and $\eta=2\ {\rm arccosh} \sqrt{5}$ and according to the above analysis $\mathfrak{S}$ is a saddle point.

Independently of $\E$, the condition $\partial_{\theta}\mathcal{P}=0$ is satisfied by the two isolated values $\theta=\pi/2$ and  $\theta=0$. These two particular values of $\theta$ can be promoted to a constant bulk solution. Indeed, they correspond to Models I and II of the previous section:
\bea
\theta=0&\qquad\Z_1=\tanh(\eta/2),& \Z_2=0\ ,\\
\nn \\ 
\theta=\pi/2&\qquad\quad\  \Z_1=\frac{1}{\sqrt{2}}\tanh(\eta/2),& \Z_2=\frac{1}{\sqrt{2}}\tanh(\eta/2)\ ,
\eea  
The existence of the space $\mathcal{M}_{\theta}$ implies that $\theta$ can be chosen arbitrarily when $\eta$ vanishes. On the other hand $\eta$ vanishes if and only if $\Z_1=0$ and $\Z_2=0$. In a superconducting solution this last condition is satisfied at the boundary of the geometry, where the space is asymptotically $AdS$. Therefore it is natural to think that black hole solutions with fixed UV value of $\theta_{\infty}\in\mathcal{M}_\theta$ can be constructed.      
In particular, superconducting solutions with $\theta_\infty\in \mathcal{M}_{\theta}$ interpolate between Model I and Model II.

We explore this possibility by analyzing the asymptotic behavior of the variables $\{\eta(r),\theta(r) \}$ in the background (\ref{Ansatz}). Expanding the equations of motion to lowest order in the $\eta$ field, we find\\  
\bea
\frac{f'}{rf}+\frac{1}{r^2}-\frac{6}{f}+\frac{\Phi'^2}{4f}=0\ ,\\
\nonumber\\
\Phi''+\left(\frac{2}{r}+\frac{h'}{h}\right)\Phi'=0\ ,\\
\nonumber\\
\E''+\left(\frac{2}{r}+\frac{f'}{f}+\frac{h'}{h}\right)\E'-\frac{1}{4}\E\ \theta'^2+\frac{\Phi^2}{4f^2h^2}\E+4\frac{\E}{f h^2}=0\ ,\\
\nonumber\\
\theta''+\left(\frac{2}{r}+\frac{f'}{f}+\frac{h'}{h}\right)\theta'+2\frac{\E'}{\E}\theta'=0\ .\label{Eqtheta2}
\eea\\
The equation for $h(r)$ is trivial, $h'(r)=0$. We suppose that the term $\theta'^2$ is sub-leading. Thus the first three equations decouple and are solved by the asymptotic expansion given in (\ref{expansF})-(\ref{expansZeta}). In particular, 
\be
\E(r)=\frac{O_{\E}^{(1)} }{r}+\frac{O_\E^{(2)} }{r^2}+\ldots\ .
\ee
Substituting into (\ref{Eqtheta2}) we find that the choice of quantization scheme for $\eta(r)$ crucially affects the asymptotics of $\theta(r)$. The equation,
\be
\theta''+\frac{4}{r}\theta'+2\frac{\E'}{\E}\theta'=0
\ee
has the desired solution only if $O_{\eta}^{(2)}=0$. Only in this case we find the admissible asymptotic behavior,  
\be
 \theta(r)=\theta_{\infty}+\frac{\xi}{r}+\ldots\ .
\ee
As $\theta'^2$ is of order $1/r^4$ we find that the assumption we made was correct. At this stage, we can think of $\xi$ as the condensate relative to $\theta_{\infty}$. In the next section we construct superconducting black holes with the following boundary conditions: $\theta_{\infty}\in\mathcal{M}_{\theta}$ and $O_{\E}^{(2)}=0$. The coefficient $O_{\E}^{(1)}\equiv O_{\E}$ define the condensate. These are the interpolating solutions that we want to construct. Our solutions offer a rich (and quite unexpected) framework where all the physical properties associated to them have a simple explanation. 

\subsection{$\Delta=1$ Condensation: Numerics at finite temperature\label{S21}}

\begin{figure}[t]
  \centering
  \subfiguretopcaptrue
  \subfigure[]{\includegraphics[scale=.5]{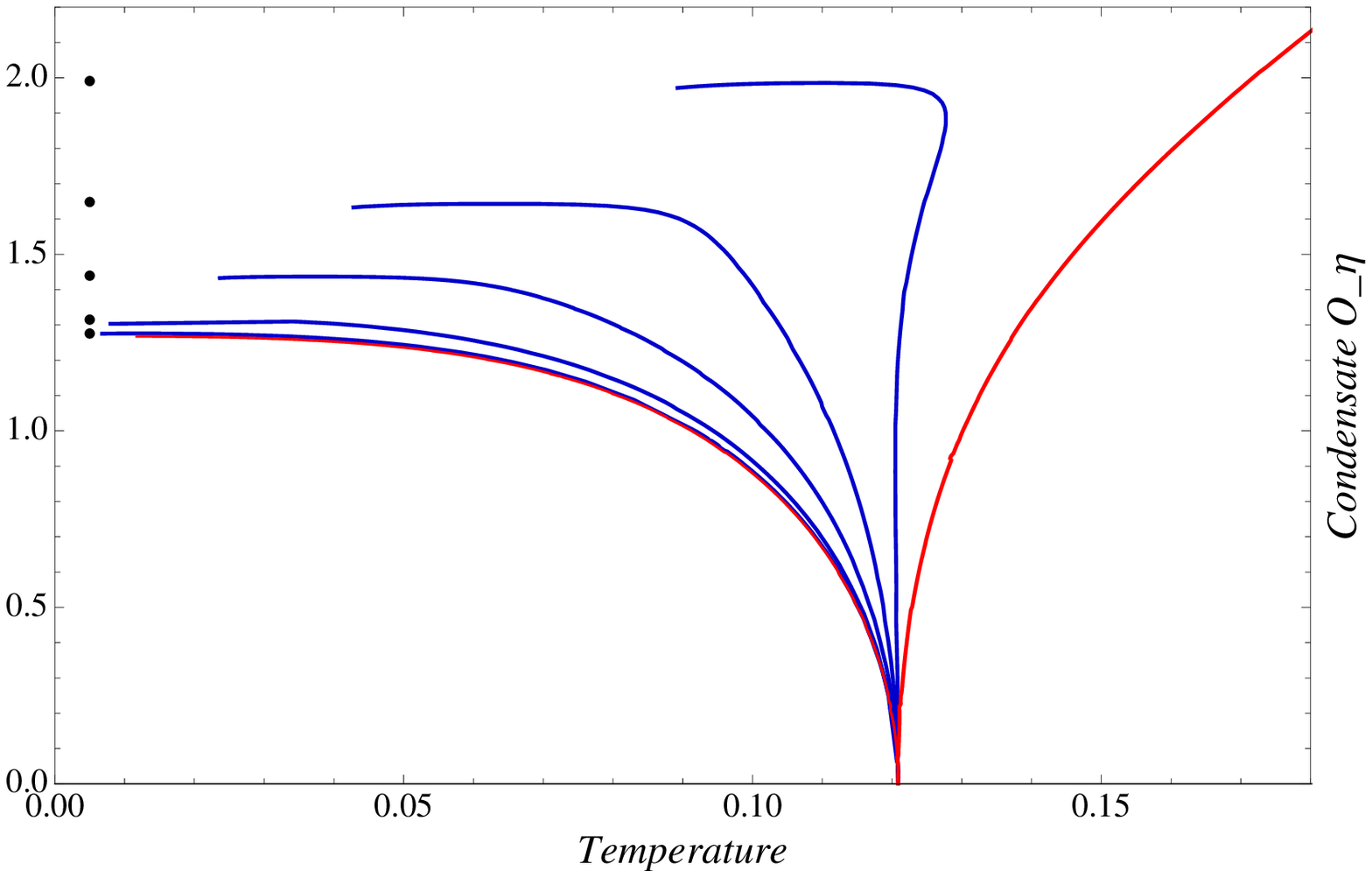}\label{Figura1A}}\hspace{1cm}
  \subfigure[]{\includegraphics[scale=.5]{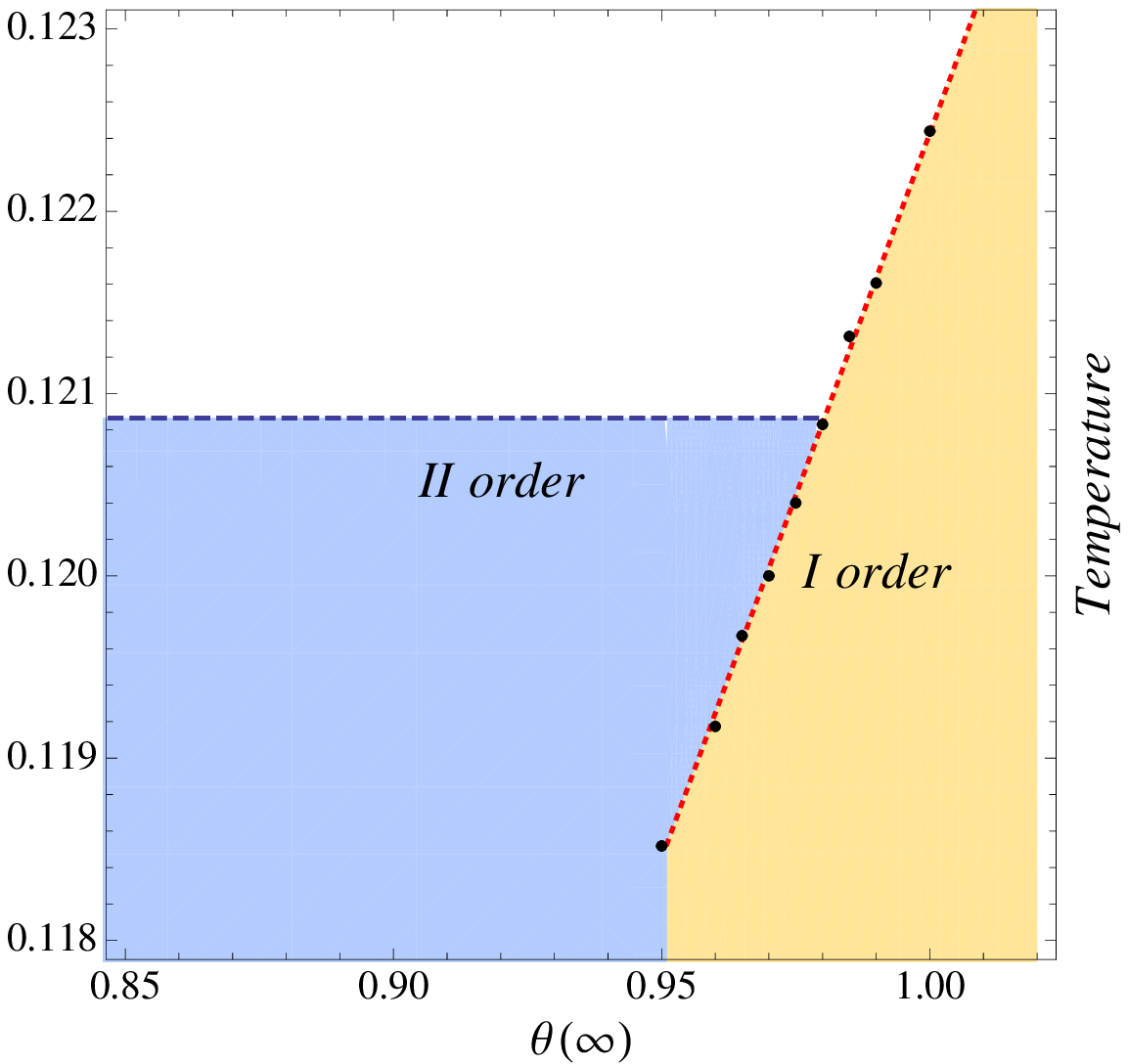}\label{Figura1B}}
  \caption{\it \small{Numerical results for finite temperature superconducting black holes. The normalization has been fixed to $\rho=1$. On the hand left side the condensate $O_{\E}(T, \theta_{\infty})$ is studied by letting the value of $\theta_{\infty}$ vary. The black dots are the values of the condensates extrapolated from our new zero temperature solutions. From bottom to top the curves correspond to $\theta_{\infty}=0,0.1,0.25,0.5,0.75,1$. In the cases $\theta_{\infty}=0$ and $\theta_{\infty}=0.1$ the zero temperature values of the condensates differ by a factor of about $0.01$. The rightmost red curve is the retrograde condensation which corresponds to $\theta_{\infty}=\pi/2$. Strictly speaking the curves should be represented in a three dimensional space according to their value of $\theta_{\infty}$. For simplicity we have collected the results in a single picture. On the right hand side the figure shows the phase diagram of the system.
}}
\vspace{-.25cm}
\end{figure}

We study the full system of equations (\ref{Eqchi})-(\ref{Eqtheta1}) with the asymptotic conditions defined in the previous section. 
The condensate ${O}_{\E}=O_{\E}(T,\theta_{\infty})$ is associated to an operator of dimension $\Delta=1$ and it is function of the temperature $T$ and of the angle $\theta_{\infty}$. Figure \ref{Figura1A} shows the temperature dependence of the condensate when the value of $\theta_{\infty}$ is varied in the range $[0,\pi/2]$.

All the curves originate from the same branch at $T\approx 0.121$. This fact is easily understood considering the linearized approximation around the Reissner-Nordstr\"om black hole \cite{Aprile:2010yb}. Indeed the critical temperature is only determined by the quadratic terms in the equation of the charged scalar $\E(r)$. In the present case all the terms depending on the field $\theta(r)$ are suppressed and therefore only the mass, $m^2L^2=-2$, and charge, $qL=1/2$, enter in the analysis. It follows that the temperature at which a small superconducting black holes exists does not depend on $\theta_{\infty}$.

Important changes happen when the value of $\theta_{\infty}$ is tuned up to the value $\pi/2$. As we expected, the shape of the curve $O_{\E}(T,\theta_{\infty})$ gets closer to the retrograde condensate, but for intermediate values, for example $\theta_{\infty}=1$, something interesting happens. In this case the shape of the curve brings to mind the case of a first order phase transition  \cite{Aprile:2009ai}. Calculating the free energy of the superconducting black holes and comparing the result with the Reissner-Nordstr\"om case, we find that the order of the phase transition depends on $\theta_{\infty}$.
The situation is clarified by Figure \ref{Figura1B}. In the white region the uncondensed solution is the only allowed configuration. When the temperature is cooled, the superconducting phase is thermodynamically favored. In the blue region the phase transition takes place at $T^{sec}_c\approx 0.121$ and it is second order. It means that the system enters the superconducting phase in a continuos way. This mean field behavior is a consequence of the large $N$ limit. In the yellow region and for temperatures above $T_c^{sec}$ the phase transition is first order. The critical temperature of the first order phase transition can be read from the contour of the yellow region: it increases along the dashed red line. In this case the curve of the condensate has two branches, the lower branch is not thermodynamically favored and when the temperature approaches the critical temperature the system jumps into the superconducting phase. The free energy shows a discontinuity similar to the cases studied in \cite{Aprile:2009ai}.

When the phase transition is still second order something new happens, even if the system has entered the superconducting phase, the free energy shows a discontinuity at some temperature smaller than $T_c^{sec}$. The new behavior starts approximately at $\theta^{crit}\approx 0.95$ and the discontinuity is of first order type. 
The discontinuity is physically interpreted as a change in the nature of the degrees of freedom that are responsible for the phase transition. Indeed, a pictorial feature of the curves $O_{\E}(T,\theta_{\infty})$ when $\theta_{\infty}$ is not small is the presence of a plateau that starts at some intermediate value of the temperature and describes the remaining part of the curve as the zero temperature is reached. The extension of the plateau grows with $\theta_{\infty}$ and when the phase transition is first order it basically represents the entire curve. In the next sections we give more details about these phenomena by calculating the entanglement entropy of the superconducting ground state. Before, we discuss the existence of the zero temperature solutions.

\subsection{Zero temperature solutions \label{SecZT}}
Figure \ref{Figura1A} indicates that for several values of $\theta_{\infty}\in\mathcal{M}_{\theta}$ the numerical black holes converge to a zero temperature solution. The case $\theta_{\infty}=0$ is well studied and for completeness we review the argument.

In a superconducting solution the value of the $\eta(r_h)$ generically runs along the potential. Small values of $\eta(r_h)$ correspond to temperatures in a neighborhood to the left of the critical temperature. In the case $\theta_{\infty}=0$, the field $\theta(r)$ vanishes in the bulk and thus the value of $\E(r_h)$ is tied to the slice $\{ \eta, \mathcal{P}(\eta,0)\}$. When the temperature is cooled, $\eta(r_h)$ hits the critical point $\mathfrak{S}$. The situation is graphically realized by the solid red line in Figure \ref{Figura0}. At the critical point there is an emergent $AdS_4$ space with radius $L^2=3/7$ and the solution of the equations of motion is,  
\be\label{IRADS}
f(r)=\frac{7}{3}r^2,\qquad h(r)=1,\qquad\eta(r)=2\ {\rm arccosh} \sqrt{5},\qquad \Phi(r)=0\ . 
\ee
Because the gauge field must carry flux at the boundary, we have to move from (\ref{IRADS}) exciting irrelevant perturbations, i.e sub-leading modes of the fields $\Phi(r)$ and $\eta(r)$. On this background the small perturbations, $\delta\eta(r)$ and $\delta\Phi(r)$, are governed by the equations
\be\nn
\delta\Phi'' + \frac{2}{r}\delta\Phi'-\frac{60}{7}\frac{\delta\Phi}{r^2}=0\ ,\qquad\qquad
\delta\eta'' + \frac{4}{r}\delta\eta'-\frac{60}{7}\frac{\delta\eta}{r^2}=0\ ,
\ee
and the solutions we need are,
\be
\eta(r)=\eta_{IR}\ r^\alpha + \ldots,\quad \Phi(r)=\Phi_{IR}\ r^\beta+\ldots,\qquad{\rm with}\qquad \alpha=\sqrt{\frac{303}{28}}-\frac{3}{2},\quad\beta=\sqrt{\frac{247}{28}}-\frac{1}{2}\ .
\ee
The extremal solution is obtained integrating the full system of equations up to the UV boundary where the shooting method is applied to match the condition $O_{\E}^{(2)}=0$. The background resembles a domain wall between the emergent $AdS_4$ and the boundary $AdS_4$ \cite{Gubser:2008wz,Gubser:2009cg}. Sometimes we refer to this solution as the conformal domain wall.
%
%

For the cases $\theta_{\infty}\neq0$ one could be suspicious about the zero temperature limit. We have already observed that the potential has a runaway direction from the saddle point $\mathfrak{S}$ towards the value $\theta=\pi/2$. Exactly for $\theta(r)=\pi/2$ the condensation is retrograde and one might think that all the curves corresponding to $\theta_{\infty}\neq0$ have to be retrograde at some point. Indeed Figure \ref{Figura0} shows that for fixed $0<\theta_{\infty}<\pi/2$ the numerics of the solutions when $T\rightarrow 0$ converges to $\theta(r_h)\rightarrow\pi/2$. In the same regime $\eta(r_h)$ grows and $\Phi'(r_h)$ takes small values. On the other hand, if the extremal limit of the finite temperature black holes exists, we should look for a solution of the equations of motion with the following behavior: $\theta(r)=\pi/2$, $\Phi(r)=0$ and divergent scalar field $\eta(r)$. Surprisingly, an analytic solution can be found and it is given by, 
\be\label{analiticT0}
ds^2=\ r^2(-dt^2+d\vec{x}^2)+\frac{dr^2}{r^2+C_{\eta}^2}\ ,\qquad \eta(r)=2\ \mathrm{arcsinh}\ \frac{C_{\eta}}{r}\ , \qquad \theta(r)=\frac{\pi}{2}\ ,\qquad \Phi(r)=0\ .
\ee
From $ds^2$ we read the definitions of the metric functions, $f(r)=r^2$ and $h(r)=\sqrt{1+C_{\eta}^2/r^2}$. We are now in the position to construct the extremal black hole. We look for irrelevant perturbations $\delta\theta(r)$ and $\delta\Phi(r)$ on the background (\ref{analiticT0}). Linearizing the equations for $\Phi(r)$ and $\theta(r)$ we find that the perturbations are governed by the equations, 
\be
\label{ZeroTPhi}
\delta\Phi''+\frac{1}{r}\delta\Phi'-\frac{2}{r^2}\delta\Phi=0\ ,\qquad\qquad
\delta\theta''+\frac{1}{r}\delta\theta'-\frac{2}{r^2}\delta\theta=0\ ,
\ee
and the solutions are,
\be\label{ZeroSolu}
\delta\Phi(r)=C_\Phi^{(1)} r + \frac{C_\Phi^{(-1)}}{r},\qquad\qquad \delta\theta(r)=C_\theta^{(1)} r+\frac{C_\theta^{(-1)}}{r}\ .
\ee 
Imposing $C_\Phi^{(-1)}=0$ and $C_\theta^{(-1)}=0$, we integrate the full system of equations, from small radial scales up to the UV boundary. There are three parameters left, $\{ C_{\eta},C_{\theta},C_{\Phi} \}$. We use the radial scaling to set $C_{\E}$ to unity and the shooting method technique to match the boundary conditions. In particular we want to fix $\theta_{\infty}$ in the range $(0,\pi/2)$ and we need to impose $O_{\eta}^{(2)}=0$. From these new zero temperature solutions we extrapolate the values of the condensates. Figure \ref{Figura1A} shows the result, the zero temperature value of the condensate, for fixed $\theta_{\infty}$, agrees with the limit $O_{\E}(T\rightarrow 0,\theta_{\infty})$ taken from the finite temperature black holes.

In the approximation $C_{\E}\rightarrow 0$ the metric in (\ref{analiticT0}) matches with $AdS_4$ and the divergent behavior of the scalar field is turned off. One might wonder if in that limit the expansion (\ref{IRADS}) is somehow recovered. We want to point out that this is not the case: for arbitrary small values of $\theta_{\infty}>0$ the limit $\theta(r\rightarrow 0)$ is always $\pi/2$. The correct interpretation of the statement is to think about the IR values, $\theta=0$ and $\theta=\pi/2$, as ``order parameters'' that distinguish two different phases of the theory\footnote{We thank R. Myers for a discussion which led to this line of thought.}. This consideration implies that the IR physics described by the conformal domain wall and the new zero temperature solution is different. Indeed, a new scale appears in the solution: $C_{\eta}$. The first geometrical feature related to $C_{\eta}$ is evident: in the region $r\ll C_{\eta}$ the metric looks like a cone in which the transverse space $M$ is the Minkowsky space,
\be
ds^2\ \approx\ dr^2+ C_{\eta}^2r^2(-dt^2+d\vec{x}^2)\ .
\ee
The cone structure can be seen explicitly through a formal construction,
\begin{itemize}
	\item For $0<C_{\eta}<1$ we define $\cos\alpha=C_{\eta}$ and we consider the following embedding. The curve $z/r=\pm\tan\alpha$ with $r\ge 0$, defines a ``surface of revolution'' in the $5$-dimensional space given by $ds^2=dz^2+dr^2+r^2dM^2$. The resulting $4$-dimensional metric is $ds^2=dr^2+r^2\cos^2\alpha\ dM^2$.\\ 
	
	\item For $C_{\eta}>1$ we define $\cosh\alpha=C_{\eta}$ and we consider a similar embedding. The curve $z/r=\pm\tanh\alpha$ with $r\ge 0$, defines a ``surface of revolution'' in the $5$-dimensional space given by $ds^2=-dz^2+dr^2+r^2dM^2$. The resulting $4$-dimensional metric is $ds^2=dr^2+r^2\cosh^2\alpha\ dM^2$.
\end{itemize}	
The construction is formal in the sense that, unless one of the transverse direction in $M$ is compact, there are no angles that can be used to rotate the curve $z=z(r)$. At $r=0$, the bulk caps off with a ``good'' conical singularity. Indeed, according to the usual criteria, the on-shell potential is bounded from above \cite{Gubser:2000nd}. We can check the statement looking at the value of the potential at origin $r=0$. In this case $\theta(0)=\pi/2$ and $\mathcal{P}(\E(r),\pi/2)=-2(2+\cosh\E)=-2(3+2C_{\E}^2/r^2)$ goes to $-\infty$ in the limit $r\rightarrow 0$. On the other hand, for intermediate scales $C_{\E}\ll r\ll \tilde{r}$, where $\tilde{r}$ is such that deviations of the bulk solution from (\ref{analiticT0}) are important, the metric still looks like $AdS_4$. 
A quantity that may be able to distinguish the differences between the conformal domain wall and the new zero temperature solution is the entanglement entropy. We dedicate the next section to this calculation.

\begin{figure}[t]
  \centering
  \subfiguretopcaptrue
  \subfigure[]{\includegraphics[scale=.45]{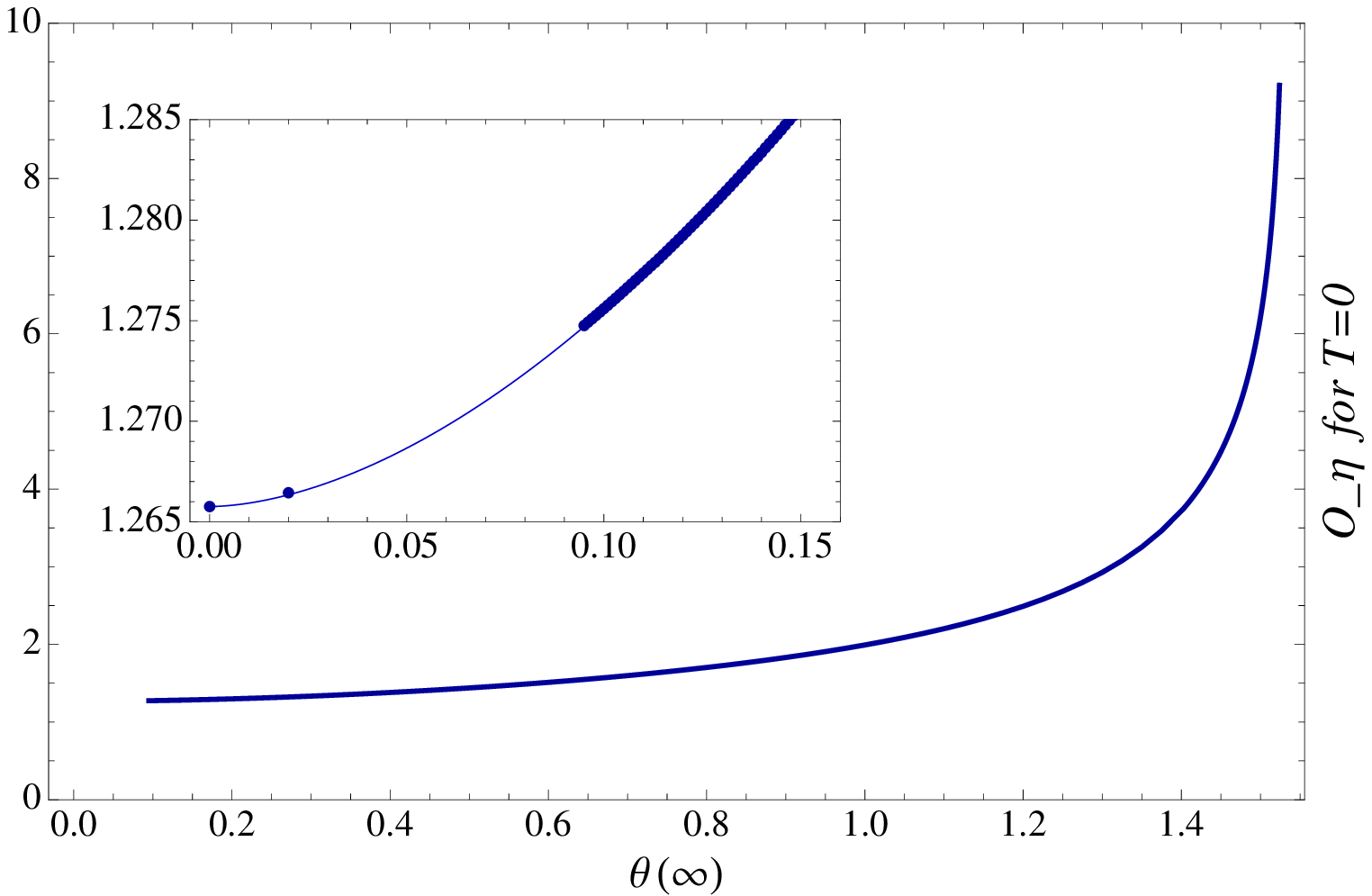}}\hspace{1cm}
  \subfigure[]{\includegraphics[scale=.45]{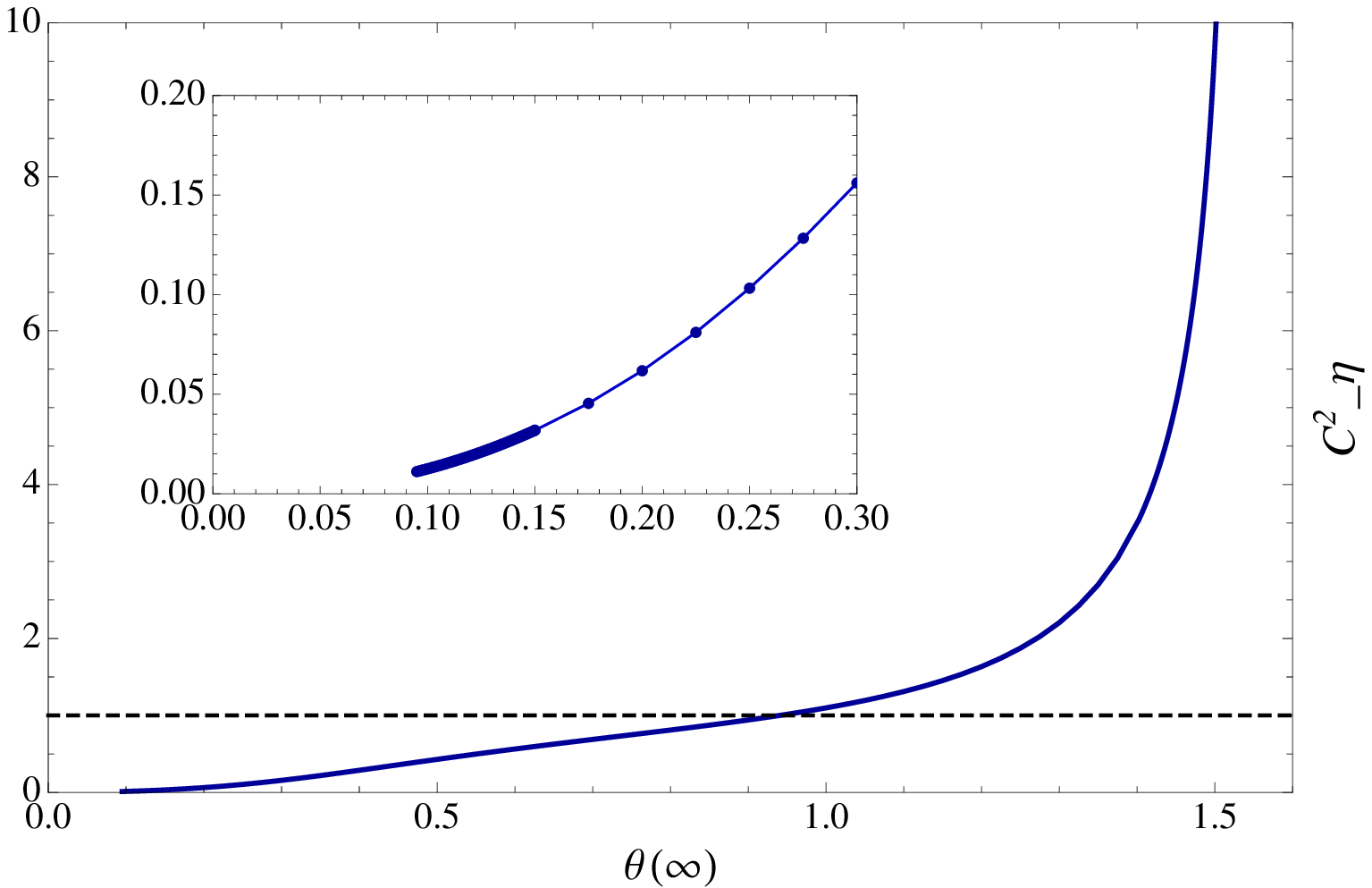}}
  \caption{\it \small{From right to left, the functions $C^2_{\E}(\theta_{\infty})$ and $O_{\eta}(T=0,\theta_{\infty})$. The zero temperature solutions are normalized taking $\rho=1$. According to our numerical precision, the best fit of the curve $O_{\eta}(T=0,\theta_{\infty})$, for small values of $\theta_{\infty}$, is $\big(O_{\eta}-1.265\big)\approx 0.55\ \theta_{\infty}^{(1.75)}$. It has been obtained by considering a log-log plot of the curve. The values of $O_{\eta}(T=0,\theta_{\infty})$ not available from the numerics of the zero temperature solutions can be obtained taking the limit of the corresponding finite temperature black holes. Even in this case the fit agrees with the numerical data. On the right hand side, the dashed black line highlights the value $C_{\E}=1$. As mentioned in this section  \ref{SecZT}, the value $C_{\E}=1$ distinguishes between the two $5$d embeddings of the IR cone geometry.}}
\label{FiguraZ}
\vspace{0 cm}
\end{figure}

Figure \ref{FiguraZ} shows the relations between $C_{\eta}$, $O_{\eta}(T=0)$ and $\theta_{\infty}$. We already know two limits of these curves. The first one is obtained for $\theta_{\infty}\rightarrow 0$. In this case the solution is close to the conformal domain wall,  we can check that $C_{\eta}$ is converging to zero whereas the condensate is approaching the expected value $O_{\eta}(T=0,\theta_{\infty}=0)\approx 1.265$. The second limit is $\theta_{\infty}\rightarrow \pi/2$. We know that the condensation is retrograde and according to the numerical data the curve has no turning point towards zero temperature. As a matter of fact, both values of $C_{\eta}$ and $O_{\eta}(T=0)$ become singular meaning that the extremal solution blows up. This behavior agrees with the expectation that the curve of the condensate exists only for temperatures above $T_c$.


The last observation about the properties of the solution regards the electric flux. The electric flux $\mathcal{F}(r)$ is given by the expression, $\mathcal{F}(r)=r^2h(r)\Phi'(r)$. In the region  $r\ll C_{\eta}$ it behaves like $\mathcal{F}(r)\approx C_{\E}C_{\Phi}r$ and therefore, in the limit $r\rightarrow 0$, the flux vanishes. From this observation we learn that the density charge $\rho$ of the field theory originates only from the charged matter in the extra coordinate. Indeed, integrating once the equation (\ref{EqPhi}) we obtain the Gauss law for the superconducting solution. It relates the $U(1)$ charge $\rho$, the electric flux at the origin and the integral of the coupling $J(\eta,\theta)$ across the bulk. Since the flux vanishes at $r=0$, the field theory charge $\rho$ originates only from the integral of $J(\eta,\theta)$.\\

\paragraph{Observation.}
The zero temperature solutions that we have constructed answer a puzzle, pointed out in \cite{Gentle:2011kv}, about the retrograde condensate. We briefly review the argument and we refer to \cite{Gentle:2011kv} for further details. The authors study families of charged solitons,  defined by various boundary conditions, in the context of $\mathcal{N}=8$ supergravity. These solutions can usually be  parametrized by their mass. In the cases in which the mass is unbounded and the soliton can get arbitrarily large, a planar limit exists. This limit corresponds to the zero temperature solution of the holographic superconductor which belongs to the specific theory under consideration and is defined by the same boundary conditions of the solitons. Model II was analyzed and it was found that the planar limit of $\Delta=1$ charged solitons is given by (\ref{analiticT0}), but the zero temperature superconductor doesn't exist because of the retrograde condensate. Now we understand that Model II is just a special case of $\theta_{\infty}\in \mathcal{M}_{\theta}$. Nevertheless, as we have seen in this section, the solution (\ref{analiticT0}) represents the IR limit of a large class of extremal black holes with each of them being labeled by $\theta_{\infty}$.

\subsection{The Entanglement entropy and the confined cohesive phase}
\begin{figure}[t]
  \centering
  \subfiguretopcaptrue
  \subfigure[]{\includegraphics[scale=.5]{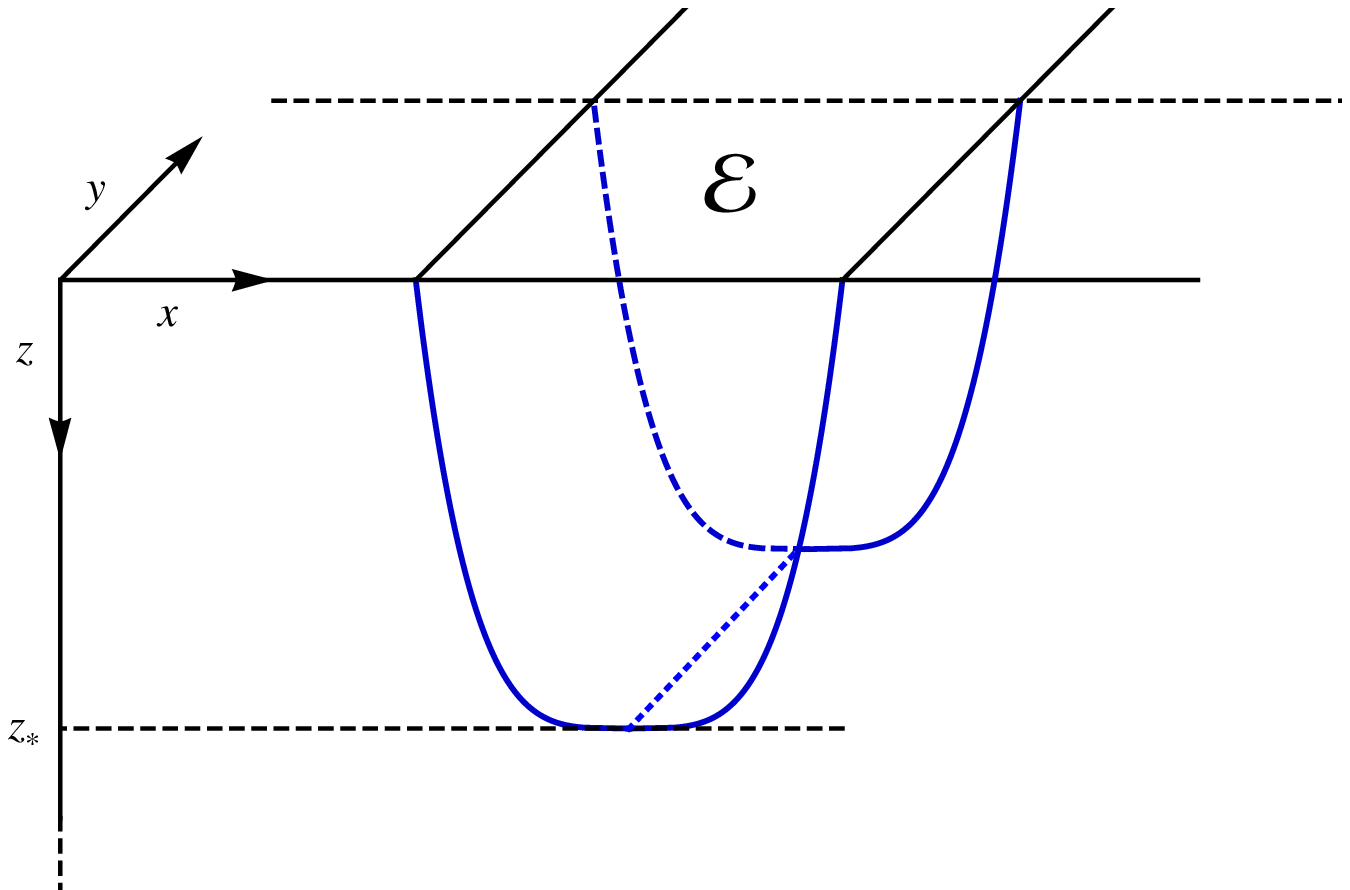}\label{FiguraEA}}\hspace{1cm}
  \subfigure[]{\includegraphics[scale=.5]{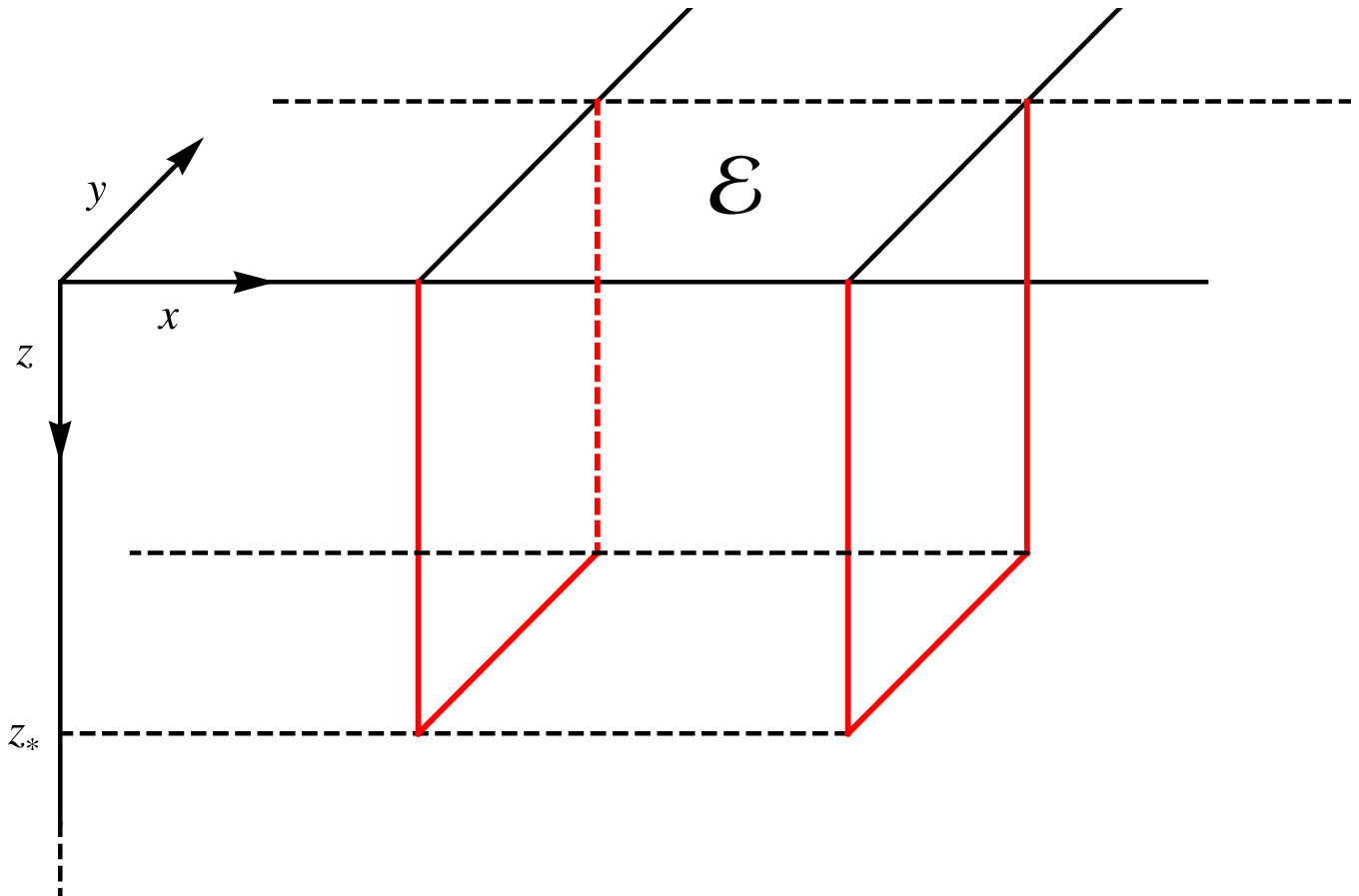}\label{FiguraEB}}
  \caption{\it \small{Cartoon of the bulk surface $\gamma_\mathcal{E}$ used in the calculation of the entanglement entropy. On the left hand side the surface is connected and the configuration has a turning point located at $z=z_\star$. On right hand side the surface is disconnected, the two planes are located at fixed $x$-coordinate and they are extended in the $z$ direction, from the boundary up to $z=z_{\star}$.
}}
\vspace{0cm}
\end{figure}

The entanglement entropy of a region $\mathcal{E}$ in the boundary field theory is computed holographically according to the proposal of \cite{Ryu:2006bv}. It is calculated by considering the area of bulk surfaces $\gamma_\mathcal{E}$ whose boundary is given by $\partial\mathcal{E}$. According to the proposal, the entanglement entropy is the minimal area, 
\be
S_\mathcal{E}=\frac{2\pi\mathrm{Area}(\gamma_\mathcal{E})}{\kappa^2}\ .
\ee
The strategy is to rewrite the problem as a variational problem. The solution of the equations of motion provides the profile of the minimal area surface and allows one to compute the entanglement entropy of the configuration. It's convenient to use the variable $z=1/r$ writing the spatial part of the metric in the form
\be
ds^2_{\mathrm{spatial}}=\frac{L^2}{z^2}\left(dx^2+dy^2+U(z)dz^2\right)\ ,\qquad U(z)=\frac{1}{z^2f(z)h^2(z)}\ .
\ee
The UV boundary is located at $z=0$. We are interested in boundary surfaces with strip shape,
\be
\mathcal{E}=\{ (x,y) |-\frac{l_x}{2}\le x\le \frac{l_x}{2},\ 0\le y\le l_y \}\ .
\ee 
We parametrize the bulk surface $\gamma_{\mathcal{E}}$ choosing coordinate $y$ and $x=x(z)$ as schematically depicted in Figure \ref{FiguraEA}. The area functional is given by,
\be\label{minimalArea}
\mathrm{Area}(\gamma_{\mathcal{E}})=2L^2\ l_y\int_\epsilon^{z_\star} \frac{dz}{z^2}\sqrt{U(z)+x'(z)^2}\ .
\ee
In the formula $z_\star$ is the turning point of the configuration whereas $\epsilon$ is an $UV$ regulator. The variational problem for $x(z)$  has a conservation law that allows to eliminate $x'(z)$ from the integral (\ref{minimalArea}). The conservation law is,
\be 
\frac{x'(z)}{z^2 \sqrt{U(z)+x'(z)^2}}=\frac{1}{z_\star^2}\ .
\ee
We solve for $x'(z)$ and we use the result to write the area functional in the form,
\be
\mathrm{Area}(\gamma_{\mathcal{E}})=2L^2\ l_y\int_0^{z_\star} \frac{dz}{z^2}\ \sqrt{\frac{U(z)}{1-(z/z_\star)^4}}\ .
\ee
Integrating the conservation law we find the relation between $l_x$ and $z_\star$,
\be\label{lunghezza}
\frac{l_x}{2}\ =\ \int_0^{\l/2}dx\ =\ \int_0^{z_\star} dz\ \frac{z^2}{z_{\star}^2}\ \sqrt{\frac{U(z)}{1-(z/z_\star)^4}}\ .
\ee
The entanglement entropy is divergent in the limit $\epsilon\rightarrow 0$. The origin of the divergence is easily understood: it corresponds to the integration of the short distances degrees of freedom and geometrically is due to the fact that the minimal surface reaches all the way to the boundary. This leading divergent term is the  ``boundary area law''  of the entanglement entropy. Subtracting the divergence, we can write $S_\mathcal{E}$ in terms of the finite quantity $s$ defined by,
\be\label{EntropiaEntanglementFinale}
S_\mathcal{E}=\frac{4\pi L^2}{\kappa^2}\ l_y\left( s+\frac{1}{\epsilon}\right)\ .
\ee
In the following the curve $s=s(\l)$ is computed using numerical calculations. Formula (\ref{EntropiaEntanglementFinale}) is proportional to the ratio $L^2/\kappa^2$, where $L$ is the $AdS$ radius and $1/\kappa^2$ is the gravitational constant of the $4$D model. According to the gauge/gravity correspondence $L^2/\kappa^2$ is a function of the numbers of colors of the dual gauge theory. The present model is not based on a string construction thus, the precise relation is not fixed. However, in the large $N$ limit the dependence $L^2/\kappa^2\sim N^{3/2}$ is to be expected \cite{Klebanov:1997kc}. The overall coefficient is not determined but the $N^{3/2}$ dependence is a robust feature.  

\paragraph{Finite Temperature.}
We begin by analyzing the results at finite temperature. A general discussion about this case has been presented in \cite{Albash:2012pd} and here we recognize basically the same features. A distinction between first order and second order phase transition in the condensate phase diagram is necessary. 

\begin{figure}[t]
  \centering
  \subfiguretopcaptrue
  \subfigure[]{\includegraphics[scale=.51]{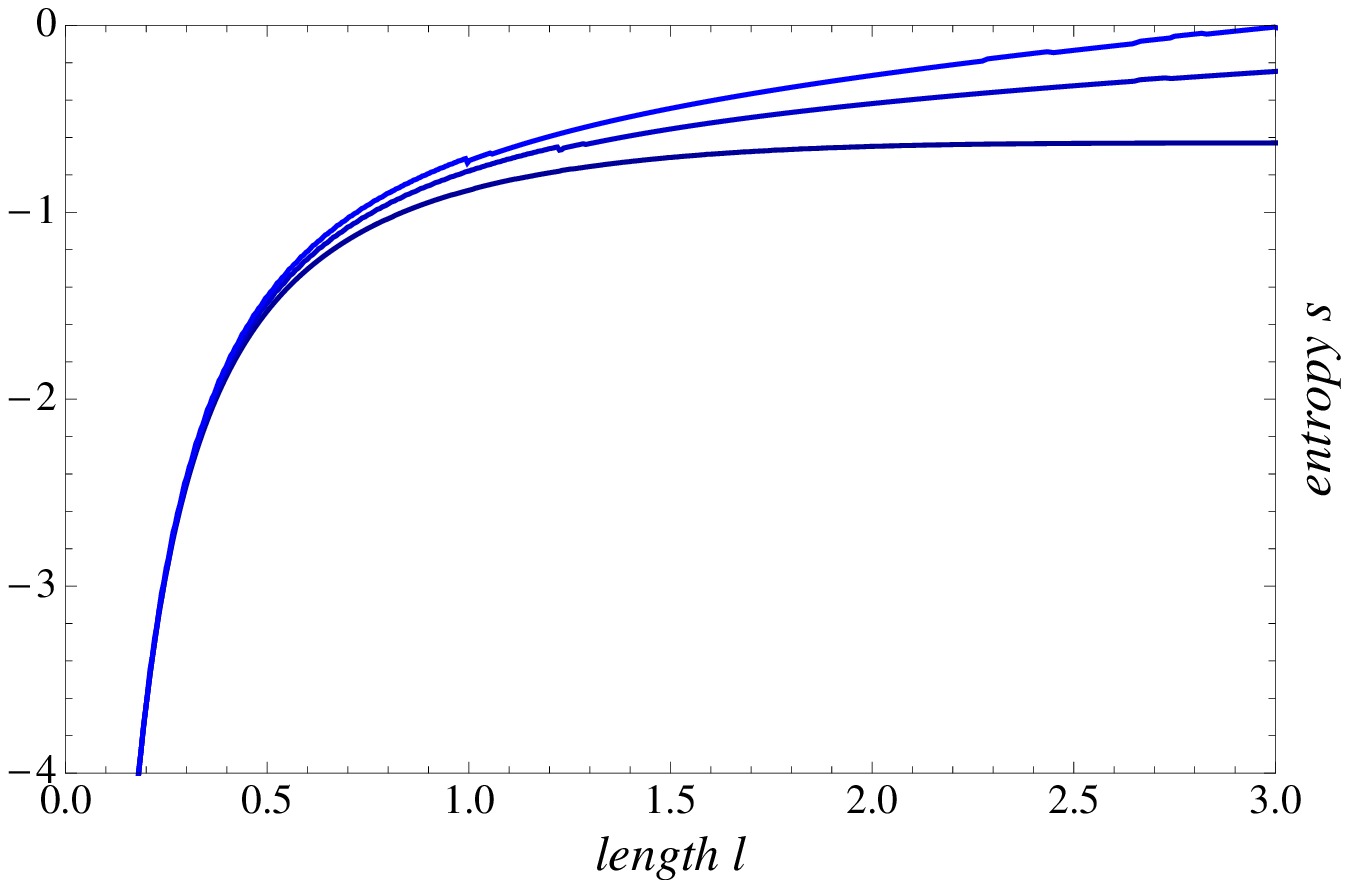}\label{Figura2A}}\hspace{1cm}
  \subfigure[]{\includegraphics[scale=.51]{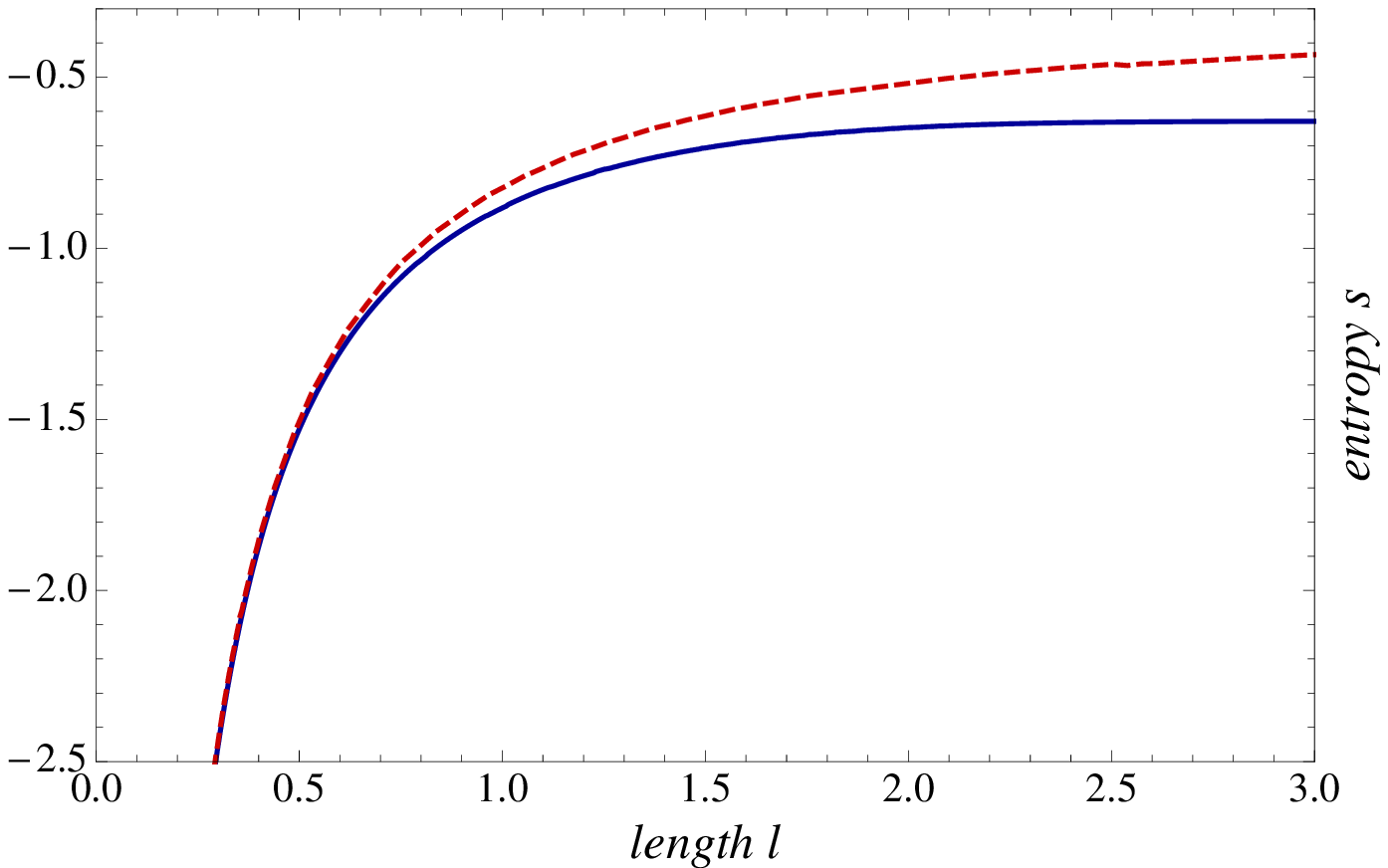}\label{Figura2B}}
  \caption{\it \small{Numerical plots of the entanglement entropy as function of the length $l_x$. On the left hand side the case $\theta_{\infty}=0.5$ is displayed for different temperatures. From bottom to top $T \approx 0.04,0.095,0.11$. On the right hand side, we show a comparison between the behavior of the entropy as function of $\theta_{\infty}$ for a fixed temperature $T \approx 0.04$. The dashed line corresponds to $\theta_{\infty}=0.1$ whereas the solid blue line is $\theta_{\infty}=0.5$. The slope for the curve corresponding to $\theta_{\infty}=0.5$ is approximately constant for large $l_x$.}}
	\label{Figura2}
\vspace{-.2cm}
\end{figure}

We first consider the case in which the phase transition is second order. In Figure \ref{Figura2} we show the plots of $s=s(\l)$ for the two values, $\theta_{\infty}=0.1$ and $\theta_{\infty}=0.5$. When the temperature is close to the critical temperature, the large lengths behavior of the entropy shows a linear dependence. Lowering the temperature, the slope drops to zero and the entropy approaches a constant value. If the region $\mathcal{E}$ has a relatively small size then $z_\star$ is close to the boundary and thus the pure $AdS_4$ result $1/\l$ is recovered.\\
In section \ref{S21} we observed that at fixed low temperature, depending on the value of $\theta_{\infty}$, the condensate $\mathcal{O}_{\eta}(T,\theta_{\infty})$ enters the plateau region. In this cases the linear behavior $s(\l)\sim \l$ drops very fast if $\theta_{\infty}$ is not small. An example is shown in Figure \ref{Figura2B}. 

\begin{figure}[t]
  \centering
  \subfiguretopcaptrue
  \subfigure[]{\includegraphics[scale=.5]{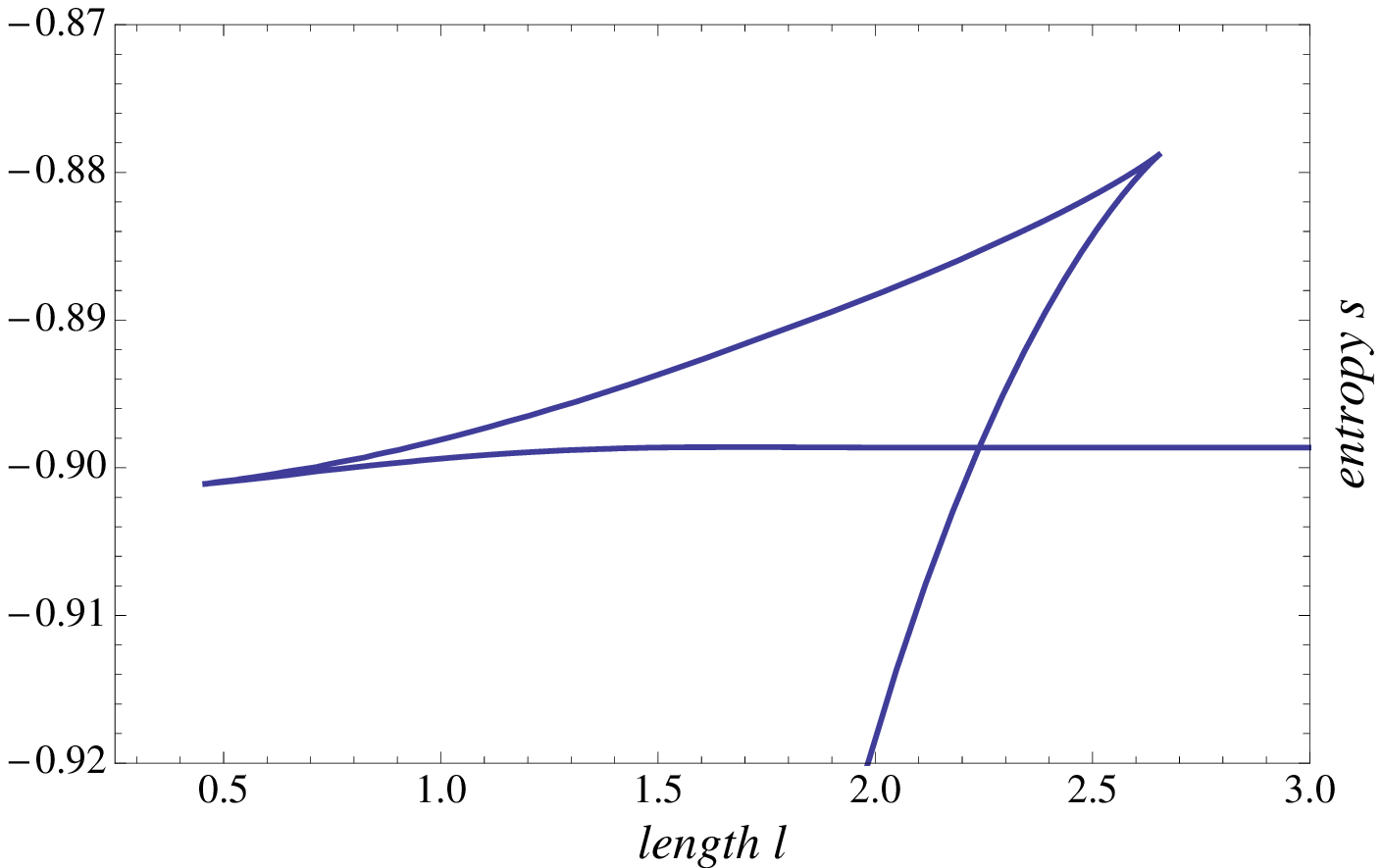}\label{Figura3A}}\hspace{1cm}
  \subfigure[]{\includegraphics[scale=.51]{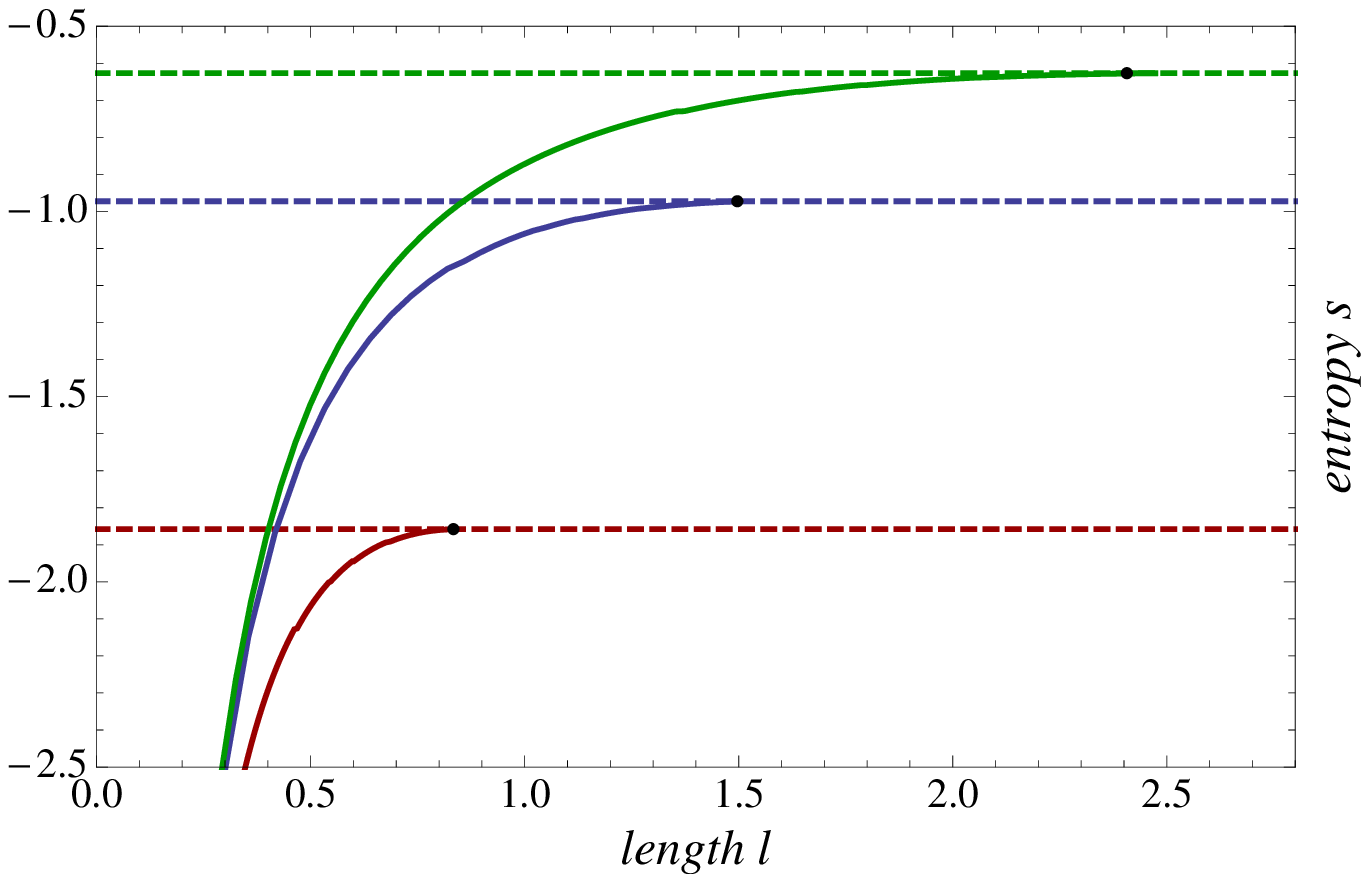}\label{Figura3B}}
  \caption{\it \small{Behavior of the entanglement entropy for several cases. On the left hand side we show a zoom of the swallowtail curve  which characterizes a first order phase transition, $\theta_{\infty}=1$ and $T\approx 0.082$.  
On the right hand side we present the result for the new zero temperature solutions. Along the solid lines the entropy has been calculated making use of the connected surface $\gamma_{\mathcal{E}}$. Dashed lines correspond to the contribution of the infinitely long disconnected surfaces. The maximum length $l^{max}$ is indicated by the black dots. From top to bottom the various cases correspond $\theta_{\infty}=0.5,1,1.4$. At $l^{max}$ the solid lines merge with the dashed lines and the confinement/deconfinement transition takes place. Interestingly, for $\theta_{\infty}=1$, $l_k$ is of the same order as $l^{max}$.}}
\vspace{-.25cm}
\end{figure}

When the phase transition becomes first order, the entanglement entropy is multi-valued at some length $l_{k}$ a kink appears. The swallowtail curve, typical of this cases, is displayed in Figure \ref{Figura3A}. For $l>l_{k}$ the slope of the curve $s(l_x)$ suddenly changes and the entropy saturates to a fixed value. 

\paragraph{Zero Temperature.} We now turn to the zero temperature solutions. For the conformal domain wall the picture does not show any substantial novelty with respect to the general analytical arguments reviewed in \cite{Albash:2011nq,Myers:2012ed}. We consider directly our new numerical results. We repeat a simple scaling argument that captures the main feature of the zero temperature solution when $\theta_{\infty}\neq 0$ \cite{Ogawa:2011bz}. 

If $z\gg 1$ the function $U(z)$ is given by the radial component of the metric in (\ref{analiticT0}),
\bea
U(z)=(1+C_\eta^2 z^2)^{-1} &\approx&  \left(C_\eta^2 z^2\right)^{-1}\qquad\  z\gg 1/C_\eta\qquad \mathrm{region\ R_1}\label{apprU}\ ,\\
&\approx & \quad 1\qquad \quad \tilde{z}  \ll z\ll 1/C_\eta\qquad \mathrm{region\ R_2} \ , 
\eea
where $\tilde{z}=1/\tilde{r}$. If the length $l_x$ is probing the intermediate region $\mathrm{R_2}$ then we expect the known behavior, however deep in the IR something new happens. We can estimate the integral (\ref{lunghezza}) by considering (\ref{apprU}) and the change of variables $z\rightarrow z/z_\star$,
\be
\frac{l_x}{2}= z_\star\int_0^{1} dz\ \frac{z^2}{\sqrt{1-z^4}}\ U(z z_\star)^{1/2}\sim const\ z_{\star}\left(\frac{1}{z_\star C_{\eta}}\right)\sim const\ \frac{1}{C_{\eta}}\ .
\ee
A maximum length $l^{max}$ exists and the above relation shows that it is proportional to $C_{\eta}^{-1}$. According to the interpretation of the radial coordinate as energy scale, $l^{\max}$ and $C_{\E}$ are correctly related. It might seem surprising that in the limit of large $z_{\star}$ the length $l_x$ remains tied to the value $l_x^{max}$. However, it does not mean that there are no configurations with $\l\ge\l^{max}$. It is important to keep in mind that the previous calculation considered only smooth and connected surfaces $\gamma_{\mathcal{E}}$, but in addition we have disconnected configurations. This class of surfaces consist of two disconnected planes located at $x=\pm l_x/2$ that are extended in the $z$ direction for a length equal to $z_{\star}$. The entanglement entropy as a function of the length $l_x$ is constant and only depends on $z_\star$ through the following formula,
\be
 S_\mathcal{E}=\frac{4\pi L^2}{\kappa^2}\ l_y\int_0^{z_\star} \frac{dz}{z^2}\ \sqrt{U(z)}=\frac{4\pi L^2}{\kappa^2} l_y\left( s+\frac{1}{\epsilon}\right)\ .
\ee
Figure \ref{Figura3B} shows that when the size of $\mathcal{E}$ is stretched up to $l^{max}_x$, there is a transition from the connected to the disconnected surface. In the limit $z_{\star}\gg 1/C_{\eta}$, the  contribution to the entanglement entropy coming from the disconnected configuration converges to the value reached by the connected surface when $l_x=l^{max}$. This transition is similar to a confinement/deconfinement transition \cite{Klebanov:2007ws}. 
The rate of change of the entanglement entropy with the length $\l$, $\partial_{l}S$, makes clear the connection with the confinement/deconfinement transition. The features of the transition are opposite from that of an Hawking-Page transition \cite{Mateos:2007ay}. For connected configurations $\partial_{l}S\sim N^{3/2}$ and the degrees of freedoms that form the ground state are in a deconfined phase. When the topology of the bulk surface changes to the disconnected configuration, the degrees of freedom living in $\mathcal{E}$ are not correlated with the ones of $\mathcal{E}^c$. This situation characterizes the large length scales and indeed the entanglement entropy is a constant for $\l>\l^{max}$: for disconnected configuration $\partial_{l}S$ vanishes. In this sense we think of $l^{max}$ as a sort of cohesion length which the entanglement entropy is able to probe. According to the classification introduced in \cite{Hartnoll:2012ux}, a confined cohesive phase is emerging in the IR of our model.

At this point we can give a more exhaustive answer to the question that we posed in the introduction: what is the effective field theory that describes the low energy properties of our zero temperature superconductors? In the case $\theta_{\infty}\neq 0$, this field theory has the characteristics of a confining theory. The geometry of the interpolating backgrounds is effectively cut off at the scale $C_{\eta}$ and a $\log$ term appears in the IR expansion of $\eta(r)$, i.e. $\eta(r)\approx\log(C_{\eta}^2/r^2)$ for $r\ll C_{\eta}$. The confined cohesive phase which emerges in the IR of our solutions is the most fascinating and intriguing novelty: the back-reaction to the presence of a mixing between the two charged hyperscalars is responsible for the confining IR geometry. On the other hand, when $\theta_{\infty}=0$, the end point of the domain wall is dual to a conformal field theory. Concerning this, it is interesting to observe that $l_x\sim 1/C_{\E}$ goes to zero in the limit $\theta_{\infty}\rightarrow 0$.

The calculation of this section helps to understand what characteristic length can be associated to the superconducting state in order to distinguish between first or second order phase transition. We consider the case $\theta_{\infty}=1$ which is particularly instructive. The phase diagram of $O_{\eta}(T,1)$ shows a clear first order phase transition. It means that correlation lengths are finite when the temperature approaches the critical temperature. Because there are only two scales in our gravitational solutions, $C_{\eta}$ and $T_c$, all other scales will be functions of $C_{\eta}$ and $T_c$. Then, we can associate the (zero temperature) value of $l^{max}$ to this first order phase transition. 

For a generic $\theta_{\infty}$ we need a more careful analysis. Indeed, we know that the maximum length $l^{max}$ exists for each value of $\theta_{\infty}$ but the phase transition is not always first order. In particular, for small values of $\theta_{\infty}$ the scale $C_{\E}\ll 1$ and therefore $l^{\max}\gg 1$. In these cases the phase transition is second order. We can make a qualitative comment about this phenomenon recalling the discussion of section \ref{S21}. When $\theta_{\infty}$ increases, the curve of the condensate develops a plateau (see Figure \ref{Figura1A}). This plateau grows until the phase transition becomes first order. At the same time $\l^{max}$ decreases towards a critical value related to $\theta^{crit}$. This critical value of the length $\l^{max}$ is the one that we associate to a ``strong'' cohesion of the ground state.

In summary, by calculating the entanglement entropy we have unfolded the interesting features codified in our extremal metric (\ref{analiticT0}). The scale $C_{\eta}$ sets the characteristic length of the confined cohesive phase. The appearance of $C_{\eta}$ in the IR of our solution has its counterpart in the fact that the superconducting state has a non trivial $\theta_{\infty}$ at the UV boundary. In the next section we investigate this connection in more detail, in particular the $AdS/CFT$ picture of our model is completed if we can describe what is the role of $\theta_{\infty}$ in the dual field theory. As we will see, the connotation of $\theta_{\infty}$ in the dual field theory is the final piece of our story.

\section{Dual field theory and marginal deformations \label{S3}}

Having studied the model in the variables $\{\eta, \theta\}$ it's convenient to go back to the original ones $\{ z_1, z_2 \}$. The asymptotic expansion for the two complex scalar fields was given in the (\ref{expansZeta}). They both have the same near-boundary behavior,
\be\label{parametri}
z_1(r)=\frac{\mathcal{O}_1^{(1)} }{r}+\frac{\mathcal{O}_1^{(2)} }{r^2}+\ldots\ ,\qquad\qquad
z_2(r)=\frac{\mathcal{O}_2^{(1)} }{r}+\frac{\mathcal{O}_2^{(2)} }{r^2}+\ldots\ .
\ee
We use the definition of $z_1$ and $z_2$ 
in terms of $\eta$ and $\theta$, i.e.  $z_1(r)=\tau \cos(\theta/2)$ and $z_2(r)=\tau \sin(\theta/2)$, to relate the coefficients $\theta_{\infty},\ \xi$ and $O_{\E}$ to the four parameters appearing in (\ref{parametri}). 
The result is the following,
\bea
\label{Coeff1}
\mathcal{O}_1^{(1)}=\frac{1}{2}O_{\eta}\cos\frac{\theta_{\infty}}{2}\ ,&\qquad &\mathcal{O}_1^{(2)}=-\frac{1}{4}O_{\eta}\xi\sin\frac{\theta_{\infty}}{2}\ ,\\
\nn\\
\label{Coeff2}
\mathcal{O}_2^{(1)}=\frac{1}{2}O_{\eta}\sin\frac{\theta_{\infty}}{2}\ ,&\qquad&\mathcal{O}_2^{(2)}=\phantom{-}\frac{1}{4}O_{\eta}\xi\cos\frac{\theta_{\infty}}{2}\ .
\eea
Given a superconducting solution with fixed $\theta_{\infty}$ none of the above coefficients is independent of the temperature. Nevertheless, defining the constant $\lambda=\tan(\theta_{\infty}/2)$, the following relations hold,
\bea
\label{RelaDT1}
	\mathcal{O}_2^{(1)}=\lambda \mathcal{O}_1^{(1)}\\
	\nn\\
\label{RelaDT2}
	\mathcal{O}_1^{(2)}=-\lambda \mathcal{O}_2^{(2)}\\
	\nn\\
\label{RelaDT3}
	\mathcal{O}_{1}^{(1)}\mathcal{O}_1^{(2)}+\mathcal{O}_2^{(1)}\mathcal{O}_2^{(2)}=0
\eea
We observe that $\lambda$ has zero dimension. From the expressions (\ref{RelaDT1}) and (\ref{RelaDT2}) we recognize all the markings of the double trace deformation in the $AdS/CFT$ setup \cite{Witten:2001ua}. In the following we explain the details of this deformation. They are important in order to understand the dual description which is behind the relations (\ref{RelaDT1}) and  (\ref{RelaDT2}). A useful technique in this context is the holographic renormalization approach. In particular, we want to evaluate the euclidean action on bulk a solution. 

Standard methods allow one to compute the euclidean action $S_E$ as a total derivative \cite{Skenderis:2002wp,Liu:2004it},
\be
S_E=-\int d^4x\ \sqrt{-g}\ \mathcal{L}\ =\ \frac{1}{2\kappa^2} \int d^3 x \int_{r_h}^{\infty} dr\ \partial_r\Big(2r h(r) f(r)\Big)=
\frac{1}{2\kappa^2}\int d^3x\ \Big(2r h(r)f(r)\Big)\Big|_{r\rightarrow\infty}\ ,
\ee
where $\mathcal{L}$ is given in (\ref{LWarner}). The surface term at the horizon vanishes both at finite temperature and at zero temperature. At finite temperature the term $r_h h(r_h)$ is finite but $f(r_h)=0$ by construction. At zero temperature the term $r h(r)$ is bounded by $C_{\E}$ in the limit $r\rightarrow 0$ but $f(r)$ vanishes like $r^2$. The surface term at infinity, i.e. in the limit $r\rightarrow\infty$, is not finite and needs to be renormalized. At the boundary we find two types of divergences,
\be\label{interm}
\Big(2r h(r)f(r)\Big)\Big|_{r\rightarrow\infty}\ =2r^3+\sum_{i=1,2}\mathcal{O}_{i}^{(1)}\mathcal{O}_i^{(1)}\ r -2M+\frac{8}{3}\sum_{i=1,2}\mathcal{O}_{i}^{(1)}\mathcal{O}_i^{(2)}\ .
\ee
The $r^3$ term originates from the integration over the $AdS$ space and it is regulated by the Gibbons-Hawking term plus a boundary cosmological constant,
\be
S_{GH}=\frac{1}{2\kappa^2}\int d^3x\ \sqrt{-g_{B}}\left(2K+\frac{4}{L}\right)\ .
\ee
The metric $g_{B}$ is the induced metric at the boundary and $K$ is the trace of the extrinsic curvature defined by
\be
K^{\mu\nu}=-\frac{1}{2}\left(\nabla^{\mu}n^{\nu}+\nabla^{\nu}n^{\mu}\right)\ ,
\ee 
with $n^{\mu}$ outward pointing unit vector, normal to the boundary.\\
The term which is linearly divergent in $r$ comes from the integration over the radial profile of the scalar fields. It can only be removed adding boundary counterterms explicitly built out of the scalars $z_1$ and $z_2$. They are not unique unless the choice of quantization scheme is fixed. In the present case, the mass of the scalar fields is $m^2L^2=-2$ and therefore both scalars can be quantized in the two possible schemes. However, there is no ambiguity. Indeed, we recall that $\Z_1$ and $\Z_2$ are part of an hypermultiplet in the $\mathcal{N}=2$ theory. Thus, their quantum numbers in $AdS$ are $\Delta=1$ and $\Delta=2$ and the natural choice is to quantize them in different ways. For concreteness we consider the alternative quantization scheme for $z_1$ and the standard quantization scheme for $z_2$. 

The scalar field $z_1$ is dual to an operator of dimension $\Delta=1$ whereas the scalar field $z_2$ is dual to an operator of dimension $\Delta=2$. The boundary values $\mathcal{O}_1^{(1)}$ and $\mathcal{O}_2^{(2)}$ are interpreted as condensates whereas, $\mathcal{O}_1^{(2)}$ and $\mathcal{O}_2^{(1)}$ are the sources. This choice is in agreement with the dynamics studied in section \ref{S21} for the value $\theta_{\infty}=0$. In this case we have set the scalar field $\Z_2$  to zero and the condensation has been associated to the operator with dimension $\Delta=1$ dual to $\Z_1$. According to this choice, the counterterm for the $z_2$ scalar is 
\be
S_2= \frac{1}{2\kappa^2} \int d^3x \sqrt{-g_{B}}\ 2\ z_2^2/L\ ,
\ee     
whereas the counterterm for the $z_1$ scalar is 
\be
S_1=-\frac{1}{2\kappa^2} \int d^3x \sqrt{-g_{B}}\left(\ 4\ z_1n^{\mu}\partial_{\mu}z_1\ +\ 2\ z_1^2/L \right)\ .
\ee
The renormalized euclidian action is finite and it is given by,
\be
\mathcal{S}_{ren}=S_E+S_{GH}+(S_1+S_2)=\frac{1}{\kappa^2}\int d^3x\ \left(-\frac{M}{2}+
\frac{4}{3}\mathcal{O}_1^{(1)}\mathcal{O}_1^{(2)}-\frac{2}{3}\mathcal{O}_2^{(1)}\mathcal{O}_2^{(2)}\right)\ .
\ee
Taking into account the relations (\ref{RelaDT1}) and (\ref{RelaDT2}), the final result is,
\be
-\mathcal{S}_{ren}=\frac{1}{\kappa^2}\int d^3x\ \left(\frac{M}{2}+ 2\lambda\mathcal{O}_1^{(1)}\mathcal{O}_2^{(2)}\right)\ .
\ee
From the above expression we understand the consequence of having identified the sources with the condensates. The double trace deformation $\mathcal{O}_1^{(1)}\mathcal{O}_2^{(2)}$, with marginal coupling $\lambda$, shows up as a finite contribution to the renormalized action. In order to be completely general, we note that the identification between sources and condensates given in (\ref{RelaDT1}) and (\ref{RelaDT2}) takes into account only the ``radial'' part of the complex fields $\Z_1$ and $\Z_2$. Indeed, looking for black hole solutions we dropped the phases from our Lagrangian however, the double trace deformation has to involve complex operators. We can easily restore the phases because the most general solution, according to our ansatz (\ref{interPAnsatz}), has the phases $\psi$ and $\varphi$ constants. Therefore we simply consider that $\Z_1= z_1(r)\mathrm{exp}({i(\varphi+\psi)/2})$ and $\Z_2= z_2(r)\mathrm{exp}({-i(\varphi-\psi)/2})$. Then, from the asymptotic expansion of $z_1(r)$ and $z_2(r)$ given in (\ref{parametri}), we obtain complex sources and complex condensates. The relations (\ref{RelaDT1}) and (\ref{RelaDT2}) are generalized taking into account the following observation: it is only consistent to identify operators with the same quantum numbers thus, the charges of the scalar fields fix the relation between sources and condensate. This means that $\Z_1$ is related to $\Z_2^{\dagger}$. Proceeding with the identification we find
\be
\mathcal{O}_2^{(1)\dagger}=\lambda_C \mathcal{O}_1^{(1)}\ ,\qquad \mathcal{O}_1^{(2)}=-\overline{\lambda}_C \mathcal{O}_2^{(2)\dagger}\ ,\qquad\mathrm{with}\quad
\lambda_C=\lambda e^{-i\psi}\ .
\ee
The marginal deformation is then,
\be\label{marGdeformation}
\delta\mathcal{S}\ =\ \int d^3x \left( \lambda_C \mathcal{O}_1\mathcal{O}_2\ +\ \overline{\lambda}_C \mathcal{O}_1^{\dagger}\mathcal{O}_2^{\dagger}\right)\ ,
\ee
where we have dropped the upper indexes and we have defined $\{\mathcal{O}_1,\mathcal{O}_2 \}$ as the complex operators dual to the scalars $\{\zeta_1,\zeta_2\}$.
Remarkably the deformation is invariant under the $U(1)$ action,
\be\label{globalsym}
\mathcal{O}_1\rightarrow e^{i\alpha}\mathcal{O}_1\ ,\qquad\qquad \mathcal{O}_2\rightarrow e^{-i\alpha}\mathcal{O}_2\ ,
\ee
and this global $U(1)$ symmetry of the field theory is not explicitly broken by $\delta S$. It is worthwhile to mention that the operator $\mathcal{O}_1^{(1)}\mathcal{O}_2^{(2)}$ is strictly marginal only in the large $N$ limit\footnote{We thank Mukund Rangamani for a remark about this point.}. 

The understanding of $\theta_{\infty}$ in the gravitational description is now clear: the parameter $\theta_{\infty}$ is mapped to the marginal coupling $\lambda$ in the dual field theory through the relation $\lambda=\tan(\theta_{\infty}/2)$. It is therefore tempting to consider the dependence on the extra coordinate $r$ of the field $\theta(r)$ in an RG fashion. In fact, it is a general feature in the $AdS/CFT$ scheme to interpret the evolution of the (zero temperature) geometry, from the $AdS$ boundary towards the bulk, as the flow of the UV microscopic theory towards a low energy regime. In this sense the interpolating solutions represent a novelty of this paper: whenever $\theta_{\infty}\neq 0$ the theory enters a low energy confining phase. However, the double trace deformation that we have identified does not break conformal invariance and therefore cannot be responsible for the RG flow. On the other hand, it is certainly evident, from the analysis of the potential in Figure \ref{Figura0}, that $\theta_{\infty}\neq 0$ drives the theory towards the $\theta=\pi/2$ well, away from the conformal fixed point represented by $\mathfrak{S}$. This is a consequence of $\mathfrak{S}$ being a saddle and cannot be avoided in the classical approximation to supergravity. Therefore, we conclude that conformal invariance is somehow broken. To approach this issue we consider the asymptotic behavior of $\theta(r)$,
$$
\theta(r)=\theta_{\infty}+\frac{\xi}{r}+\ldots\ .
$$
In a superconducting solution we have seen that $\xi\equiv 0$ implies $\theta(r)$ constant in the bulk. The value $\theta_{\infty}=0$ belongs to this case and even if it represents a trivial example, because $\lambda=0$, it matches the expectation that conformal invariance is not explicitly broken by the deformation. Therefore, $\xi\neq 0$ parametrizes our ignorance of how conformal invariance is broken in the UV \cite{Gubser:1999pk}. This observation suggests that a relevant deformation is turned on at a certain high energy scale. The same relevant deformation is then responsible for the RG flow. 
The intuition on $\xi\neq 0$ can be further specified by considering another important relation which holds between the two condensates: 
\be\label{relxi}
\mathcal{O}_2^{(2)}=\frac{1}{2}\xi\mathcal{O}_1^{(1)}\ .
\ee
This is an ``on-shell'' relation which is not specified by the boundary data given in (\ref{RelaDT1}) and (\ref{RelaDT2}). Instead, it arises from the bulk dynamics of the interpolating solution, in particular from (\ref{Coeff1}) and (\ref{Coeff2}). Because we want to keep the $U(1)$ invariance manifest, it is convenient to work with the complex version of (\ref{relxi}), 
\be\label{relxiC}
\mathcal{O}_2^{(2)\dagger}=\frac{1}{2}\xi\mathcal{O}_1^{(1)}e^{-i\psi}\ .
\ee
According to our interpretation of $\delta\mathcal{S}$ as double trace deformation, the coefficient $\xi$ assumes the role of an energy scale proportional to the beta function of the coupling $\lambda$, i.e. $ r\theta'(r)\propto \xi/r$. We conclude that the relation (\ref{relxiC}), which introduces the UV scale $\xi$ in the definition of the condensates, breaks conformal invariance and provides the relevant deformation which initiates the RG flow. Indeed, by substituting the above relation (\ref{relxiC}) in $\delta\mathcal{S}$, we find an effective UV action of the form, 
\be\label{proposal}
\delta\mathcal{S}_{eff}\ \propto\ \lambda\xi\mathcal{O}_1^{\dagger}\mathcal{O}_1\ .
\ee
This effective action has the expected RG behavior and in fact, by using dimensional analysis, we find that $\int d^3x\ \delta\mathcal{S}_{eff}\propto (\xi/E)$. Once the RG flow is initiated the coupling $\lambda$ runs. Its value in the IR, given by $\theta(0)=\pi/2$, characterizes the effective low energy confining phase with respect to the conformal point identified by $\mathfrak{S}$. In particular, the field theory dynamics of the tachyonic direction at the saddle point can be explored with the following argument. The coordinates $\Z_1$ and $\Z_2$ also diagonalize the Hessian in a neighborhood of $\mathfrak{S}$. The $\Z_2$ direction corresponds to a scalar with the negative mass $m^2L^2=-12/7$\footnote{Note that the mass value is above the BF bound and therefore $\mathfrak{S}$ is a stable non supersymmetric fixed point \cite{Fischbacher:2010ec}.}, dual to a relevant operator in the IR conformal theory. Then, any mixing between $\Z_1$ and $\Z_2$ in the UV will source this IR relevant perturbation. In other words, even if $\xi$ is of order $\epsilon$, and the theory stays arbitrarily close to the conformal point $\mathfrak{S}$ at an intermediate scale, the mentioned IR relevant perturbation will drive the flow along the tachyonic direction. The above intermediate scale is $C_{\eta}$. This fact can be seen both from the radial component of the metric in (\ref{analiticT0}) and from the $\log$ term that appears in the IR expansion of $\eta(r)$, i.e. $\eta(r)\approx\log(C_{\eta}^2/r^2)$ for $r\ll C_{\eta}$.   

Finally, the deformation $\delta\mathcal{S}$ is exactly marginal only for the two cases $\theta_{\infty}=0$ and $\theta_{\infty}=\pi/2$. It is therefore interesting to note that when $\theta_{\infty}\neq 0$ the coupling flows along the flat direction towards $\theta=\pi/2$. If $\theta=\pi/2$ was a second fixed point, this kind of flow would be the expected running of an exactly marginal operator when some massive field is integrated out \cite{Leigh:1995ep}. It is a nice property of our model that $\theta(0)=\pi/2$ still characterizes the IR effective theory even if it does not correspond to any fixed point. In this sense the supergravity interpretation, that  $\theta_{\infty}=0$ and $\theta_{\infty}=\pi/2$ are better understood as ``order parameters'' for the different phases of the model, is also valid in the dual field theory. In the next section, we seriously take into account the properties of $\delta\mathcal{S}$ 
by considering how our model satisfies some general statement about Landau-Ginzburg theories. 

Despite the remarkable interpretation of our extremal interpolating solutions in terms of RG flow dynamics, a field theoretical argument able to explain the reason why the retrograde condensate does not loop back to zero temperature is still absent. It remains an open problem and unfortunately our analysis of Section \ref{SecZT} only provides other evidence for 
the non existence of a zero temperature solution associated to the retrograde condensate.
  
\subsection{Thinking about the dual field theory}
We want to suggest a feasible connection between the existence of the moduli space $\mathcal{M}_{\theta}$ and the existence of a marginal operator in the dual field theory. We borrow part of the story from the theory of the $\mathcal{N}=2$ Landau-Ginzurg models in two dimensions. Reviewing these ideas we closely follow \cite{FRE}. 

We recall the notion of moduli in the Landau-Ginzburg model. A pedagogical example, which is also useful to understand the nature of phase transitions, is the mean field theory of a single scalar field $\varphi$. The potential of the theory is given by
\be\label{potLandau}
V=m^2(T)\varphi^2+\lambda\varphi^4\ .
\ee
with $T$ a continuous parameter (the temperature) and with $\lambda$ constant, positive and greater then zero. In a stable configuration the scalar is seated at the minimum of the potential. A smooth phase transition is realized if the parameter $m^2(T)$ takes negative values in some range of temperatures. If $m^2(T)$ behaves as follow,
\be
\left\{\begin{array}{ccc}
	m^2(T)<0&\qquad &\mathrm{for\ }T<T_c\ ,\\
	m^2(T)=0&\qquad &\mathrm{for\ }T=T_c\ ,\\
	m^2(T)>0&\qquad &\mathrm{for\ }T>T_c\ ,\\
\end{array}\right.
\ee
the extrema of the potential, which is a solution of the equation,
\be\label{minimumsol}
\partial_{\varphi} V(T,\varphi)\Big|_{\varphi=\varphi_0}=0\ ,
\ee
changes with the temperature. Above the critical temperature the potential is a sum of positive quantities and the only solution to (\ref{minimumsol})
is $\varphi_0=0$. Below the critical temperature the value $\varphi_0=0$ becomes a local maximum and the new minimum, 
\be
\varphi_0=\sqrt{\frac{-m^2(T)}{\lambda}}\ ,
\ee
is the stable configuration. We want to emphasize a feature of $V(T_c)$ that is physically crucial. At the critical temperature the equation
\be
\partial_{\varphi} V(T_c,\varphi)\Big|_{\varphi_0}=\lambda \varphi^3_0=0
\ee
has solution $\varphi_0=0$ which is three times degenerate: we say that $V(T_C,\varphi)$ is critical. Starting from this example we analyze a more general situation. The definition of a critical potential falls into the framework of {\it singularity} functions. By considering $n$ field variables, we define a (polynomial) potential $\mathcal{V}(X_1,\ldots,X_n)$ to be critical, or a {\it singularity} function, if its critical points are degenerate.\\ 
The possible deformations of $\mathcal{V}(X_1,\ldots,X_n)$ are polynomials in the original field variables and are classified according to the renormalization group as: relevant, marginal and irrelevant. Relevant perturbations split the degeneracy, but marginal deformation do not. The presence of marginal operators reveals that the critical potential is not isolated, but rather it is an element of a family $\mathcal{M}_V$ of functions. If $k$ is the number of marginal operators then the elements of $\mathcal{M}_V$ are labeled by $k$ continuous parameters $\{\lambda_1,\ldots, \lambda_k\}$ which are the coefficients of the marginal perturbations. In other words, an element of $\mathcal{M}_V$ is of the form,
\be
\mathcal{V}(X_1,\ldots,X_n,\lambda_1,\ldots,\lambda_k)=\mathcal{V}(X_1,\ldots,X_n)+\sum_i^k\ \lambda_i\ \mathcal{F}_i\ ,
\ee
where $\mathcal{F}_i$ is a marginal operator. The space $\mathcal{M}_V$ has the structure of a ring and the coefficients $\{\lambda_1,\ldots, \lambda_k\}$ are called moduli. Each potential living in $\mathcal{M}_V$ admits a critical point which is degenerate. 

If we want to relate the degeneracy of the potentials in $\mathcal{M}_V$ to our theory, Figure \ref{Figura1A} clearly shows that all the condensates arises from the same branch at $T\approx 0.121$, independently of $\theta_{\infty}$. We can say more about the relation between the spaces $\mathcal{M}_V$ and $\mathcal{M}_{\theta}$. Taking into account the results of the previous section we can collect the following chain of observations. First, the double trace operator $\mathcal{F}=\mathcal{O}_1\mathcal{O}_2$ is a marginal operator which belongs to $\mathcal{M}_V$ in the dual field theory. Second, the modulus $\lambda$ associated to $\mathcal{F}$ is geometrically the parameter $\theta_{\infty}$ which belongs to $\mathcal{M}_{\theta}$. Thus, the $AdS/CFT$ correspondence maps the modulus  associated to $\mathcal{F}$ to the space $\mathcal{M}_{\theta}$. More precisely, the relation $\lambda_C=\tan(\theta_\infty/2)e^{-i\psi}$, together with the restriction $\theta_{\infty}\in [0,\pi/2]$, which defines $\mathcal{M}_{\theta}$, implies that $\lambda_C$ parametrizes the unit ball on the complex plane. We remove from this set the circle $S^1$ because of the retrograde condensate and we refer to the open ball as $\mathcal{B}(0,1)$. At this point we can rephrase the analysis of the previous section by considering the action of the renormalization group flow on the open ball. We know that the origin is a fixed point and therefore we focus on the set  $\mathcal{B}(0,1)\setminus\{ 0\}$. In this case the renormalization group flow acts as an projection and the IR image of $\mathcal{B}(0,1)\setminus\{ 0\}$ is the unit circle.\\
At the classical level one might expect $\lambda_C$ to vary in the entire complex plane. Instead, we find that $\lambda_C\in\mathcal{B}(0,1)$. It is natural to ask what happens to the complement of $\mathcal{B}(0,1)$. In this case $\theta_{\infty}$ takes values in the range $[\pi/2,\pi]$. We already know the dynamics of the model simply because the potential is $\pi/2$-periodic: when $\theta_{\infty}$ is increased from $\pi/2$ to $\pi$, the condensate goes backwards from the retrograde condensate to the conformal domain wall. These solutions are related to the ones found for $\theta_{\infty}\in\mathcal{M}_{\theta}$ but they are not the same. In fact, for $\theta_{\infty}=\pi$ the scalar field $\Z_1$ is set to zero and the condensation is driven by $\Z_2$. This situation is the opposite of the case $\theta_{\infty}=0$. Let's see what is the boundary description in this case. The relations 
\be\label{Altre}
\mathcal{O}_2^{(1)}=\lambda \mathcal{O}_1^{(1)}\ ,\qquad 	\mathcal{O}_1^{(2)}=-\lambda \mathcal{O}_2^{(2)}\ ,
\ee
are always valid, even in the range $\theta_{\infty}\in [\pi/2,\pi]$. In the limit $\theta_{\infty}\rightarrow\pi$, the coupling $\lambda$ blows up and the relations (\ref{Altre}) make sense only if $\mathcal{O}_1^{(1)}\rightarrow 0$ and $\mathcal{O}_2^{(2)}\rightarrow 0$. These two conditions are suitable for the opposite quantization scheme from that adopted in $\mathcal{B}(0,1)$: the condensates are now $\mathcal{O}_2^{(1)}$ and $\mathcal{O}_1^{(2)}$, the sources instead are $\mathcal{O}_1^{(1)}$ and $\mathcal{O}_2^{(2)}$. Thus, the conformal domain wall in the case $\theta_{\infty}=\pi$ has to be associated with the condensation of the operator $\mathcal{O}_2^{(1)}$ with $\mathcal{O}_2^{(2)}=0$. In summary, the quantization schemes in the region $\mathbb{C}\setminus\overline{\mathcal{B}(0,1)}$ is opposite from that in $\mathcal{B}(0,1)$.

\section{Conclusions and Outlook}

In this paper we have studied a particular $\mathcal{N}=2$ supergravity theory in four dimensions. We based our analysis on the general principles of the $AdS/CFT$ correspondence focusing our attention on the holographic superconductivity framework. The theory has interesting dynamics and several new ingredients coexist in the final picture. These are, the interpolating solutions constructed in section $2$, the properties of the zero temperature solutions outlined in section $3$ and the notion of double trace deformation in $AdS/CFT$. Each of them can be related to the existence of the space $\mathcal{M}_{\theta}$ which is completely tied to the nature of the hypermatter scalar manifold $SU(2,1)/U(2)$. We emphasize the topological properties of the manifold $SU(2,1)/U(2)$ in the construction of the model. The manifold $SU(2,1)/U(2)$ is homeomorphic to a ball in $\mathbb{C}^2$ and the coordinates (\ref{interPAnsatz}), for fixed $\tau\neq 0$, are indeed related to the Hopf fibration of the three-sphere. In this description the angle $\theta$ is the azimuthal angle of the two-sphere whereas the polar angle $\varphi$ and the fiber $\psi$ are phases. Phases cannot appear in the potential and the potential $\mathcal{P}$ is function only of $\tau$ and $\theta$. At the origin, $\tau=0$, the compact space shrinks. This fact manifests itself in a ``topological'' degeneracy and $\mathcal{M}_{\theta}$ shows up as a moduli space.

$\mathcal{N}=2$ supergravity puts strong constraints on the nature of the scalar manifold: hypermultiplets parametrize quaternionic K\"ahler manifolds and vector multiplets parametrize special K\"ahler manifolds. For the quaternionic case, the most relevant manifolds are the homogeneous spaces of real dimension $4m$ \cite{deWit:2001dj,VanProeyen:2001wr},
\be
X(m)=\frac{SU(2,m)}{SU(m)\times SU(2)\times U(1)}\ .
\ee
The lowest dimensional space corresponds to $m=1$ and it is the coset manifold $SU(2,1)/U(2)$. In this sense we can say that the hypermultiplet $\{\Z_1,\Z_2\}$, considered as building block of our model, is universal. Then, the graviphoton together with the metric connections, are the minimal elements needed to construct the Lagrangian (\ref{LWarner}). It may be possible that the topology of higher dimensional quaternionic manifold leads to a more general moduli space. In this case the gauging of the $U(1)$ isometries requires the presence of vector multiplets. In four dimensions vector multiplets contain complex scalars and the interplay among all the matter fields can be quite involved. If the deformations $\mathcal{F}_i\in\mathcal{M}_V$ are constrained by some $U(1)$ global invariance, then the relevant gauging can only pick the $\sigma_3$ direction in the $SU(2)$ part of the isotropy group. For that reason, candidates to describe a more general moduli space are scalar manifolds which contain copies of $SU(2)$ in their isotropy group. On the other hand, the presence of several gauge fields in the gravitational action introduces more than one $U(1)$ charge in the dual field theory. The complexity of the bulk system is therefore justified by the complicated dynamics of the field theory.

The class of 5D $\mathcal{N}=2$ supergravity theories coupled to matter fields is also very well understood. In this context, the gauging of the universal hypermultiplet follows the same procedure outlined in section \ref{S1}. When the $U(1)$ killing vector is taken to be the $\sigma_3$ direction in the $SU(2)$ part of the isotropy group, the two complex scalar fields parametrizing the quaternionic manifold have opposite charges and the same mass. In particular, the mass lies at the conformal value $m^2L^2=-15/4$. The model has been proposed as a truncation of the type IIB theory on $AdS_5\times T^{1,1}$ and the two scalars might describe the dynamics of the chiral operators $\OO_{ij}=Tr[A_iB_j]$ in the dual field theory \cite{Klebanov:1998hh}. In the next paragraph we make a more precise comment about the implications of having a brane construction of a generic model. Concerning this 5D dimensional truncation, we only mention that the model contains a standard holographic superconductor and a retrograde condensate \cite{Aprile:2012ai}. In principle, the analysis carried out in the present paper has a straightforward generalization. Indeed, an extremal solution similar to (\ref{analiticT0}) exists also in this five dimensional case and is given by,
\be
f(r)=r^2\ ,\quad h(r)=\sqrt{1+\frac{C_{\eta}^3}{r^3}}\ ,\quad \eta(r)=2\ \mathrm{arcsinh}\ \left(\frac{C_{\eta}}{r}\right)^{3/2}\ ,\quad \Phi(r)=0\ ,\quad
\theta(r)=\frac{\pi}{2}\ .
\ee
However, the radial component of the metric, $g_{rr}=(f(r)h^2(r))^{-1}$, goes to zero like $r$ in the limit $r\rightarrow 0$ and therefore the IR geometry is not a cone. It is not clear whether this behavior could affect substantially the physics of the model.  

The Lagrangian (\ref{LWarner}) belongs to the category of the ``top-down'' models. With this name it has been indicated the set of all those theories that can be constructed by first principles. In general, a top-down model is a model which admits a description in terms of branes, either in the type IIB theory or in the M-theory. Having a brane description of the theory allows a more precise understanding of the relation between bulk fields and boundary operator. This statement has a strong physical implication, it means that the theory has a microscopic description in terms of open strings. Thus, the potential and other couplings of the model are fixed by the string dynamics. Several examples have been found and discussed \cite{Gubser:2009qm,Gauntlett:2009dn,Arean:2010wu,Gauntlett:2009bh}.
In our case the Lagrangian (\ref{LWarner}) can be embedded in a consistent truncation of the four dimensional $\mathcal{N}=8$ supergravity however, the relation with the ABJM theory encounters a spontaneous obstacle. Indeed, a remark in \cite{Bobev:2011rv} shows that the boundary conditions for a $U(1)$ spontaneous symmetry breaking superconductor, allowed by this theory, are restricted. In particular, if we insist on the spontaneous symmetry breaking, the microscopic theory for our boundary conditions (\ref{RelaDT1}) and (\ref{RelaDT2}), is not known. The present paper provides evidence of a beautiful connection between the scalar manifold of the gravitational theory and the space of marginal deformations of the dual field theory. If the dual field theory admits a brane construction it might be possible to provide a general statement about this type of connection. In this case the supergravity Lagrangian is obtained by KK reduction and the scalars of model parametrize the complex structure of the compact internal space, on the other hand this compact internal space has to do with the moduli space of the field theory. These kind of ideas deserve further investigation.    

Finally, we want to stress the importance of the retrograde condensate. It lives at the border of the moduli space and even if it is not relevant to the thermodynamics of the system, it does signal the existence of more general solutions. Furthermore $\theta=\pi/2$ is also the IR limit of the renormalization group flow of the marginal coupling constant and the value for which the charged soliton and the extremal solution (\ref{analiticT0}) have been found. Another example of retrograde condensate arises in the context of $\mathcal{N}=8$ supergravity in five dimensions whose dual field theory is $\mathcal{N}=4$ SYM \cite{Aprile:2011uq}. The sector of the $\mathcal{N}=8$ theory in which the retrograde condensate has been found is the $\mathcal{N}=2$ Lagrangian of the STU model coupled to three hypermultiplets \cite{Liu:2007rv}. We suggest that the theory could have an interesting structure not yet explored. 
The study of this case would be particularly relevant for understanding the phase diagram of $\mathcal{N}=4$ SYM. 

\section*{Acknowledgements}

It is a pleasure to acknowledge helpful discussions with Rob Myers, Mukund Rangamani and Matteo Bertolini. We thank Jorge Russo for bringing the model (\ref{LWarner}) to our attention. We thank Jorge Russo, Diederik Roest, Andrea Borghese and Aldo Dector for useful conversations and remarks about the present work and related topics. F.A. is grateful to Perimeter Institute for hospitality. F.A. is supported by a MEC FPU Grant AP2008-04553. This research was supported in part by Perimeter Institute for Theoretical Physics. Research at Perimeter Institute is supported by the Government of Canada through Industry Canada and by the Province of Ontario through the Ministry of Economic Development \& Innovation.

\appendix
\section{Black hole solutions: horizon expansion and thermodynamics}
The superconducting black holes are obtained integrating the equation of motions from a postulated horizon. It is a numerical problem and in order to be well posed we have to discuss how the field variables are expanded around this horizon. We use the variables $\{f(r),h(r),\Phi(r),\eta(r), \theta(r)\}$ that are the ones in which the black hole solutions have been constructed (see section \ref{S21}). The general strategy to attack the problem is explained in \cite{Hartnoll:2008kx}, here we repeat the main ingredients. 

A regular event horizon is defined by the greatest simple root of the metric components, we call this root $r_h$. The temperature of the black hole is the Hawking temperature,
\be\label{Hawking}
T=\frac{1}{4\pi}\ \sqrt{f'(r)\big(f(r)h^2(r)\big)'\ }\ \Big|_{r=r_h}\ .
\ee 
The metric component $g_{tt}$ and $g_{rr}$ have $r_h$ as a common zero. For the numerical problem we can choose the gauge
\be\label{horizonUno}
f(r)=f'(r_h)(r-r_h)+\ldots\ ,\qquad h(r)=h(r_h)+h'(r_h)(r-r_h)+\ldots\ .
\ee 
in which $f(r_h)=0$ but $h(r_h)$ is finite. Being the horizon a simple root we need to consider $f'(r_h)\neq 0$. The equation of motion for the Maxwell field $\Phi$ involves the time component of the contravariant quantity $A^{\mu}$. For $A^{t}$ to be well defined we require $\Phi(r)$ to vanish like $f(r)$,
\be
\Phi(r)=\Phi'(r_h)(r-r_h)+\ldots\ .
\ee
Then the equation (\ref{EqPhi}) is well defined at the horizon. The two scalar fields, $\eta(r)$ and $\theta(r)$, are assumed to be constant at the horizon, therefore their expansion is
\be
\eta(r)=\eta(r_h)+\eta'(r_h)(r-r_h)+\ldots\ ,\qquad\theta(r)=\theta(r_h)+\theta'(r_h)(r-r_h)+\ldots\ .
\ee
In this case the equation of motion (\ref{Eqpsi}) and (\ref{Eqtheta1}) are well defined only if the divergent term, proportional to $1/f(r)$, cancels. This requirement imposes a condition on the value of the first derivatives of the scalars field. In this case, $\eta'(r_h)$ and $\theta'(r_h)$ are given by,
\be
\eta'(r_h)=\frac{\partial_{\eta}V}{f' h^2}\Big|_{r=r_h}\ ,\qquad \theta'(r_h)=\frac{1}{\sinh^2\left(\E/2\right)}\frac{\partial_{\theta}V}{f' h^2}\Big|_{r=r_h}\ .
\ee
The coefficient $f'(r_h)$ is not an independent quantity, in fact its value is fixed by the consistency of the equation (\ref{Eqg}),
\be
f'(r_h)=-\frac{r_h}{2 h^2}\left(\mathcal{P}+\frac{1}{2}\Phi' h^2\right)\Big|_{r=r_h}\ .
\ee
Finding the solution to the equations of motion is now a well posed Cauchy problem. We can list how many free parameter are left, 
\be 
r_h\ ,\qquad h(r_h)\ ,\qquad\Phi'(r_h)\ ,\qquad\ \eta(r_h)\ ,\qquad \theta(r_h)\ .
\ee
This set is further reduced by considering that the equations of motion are invariant under the two scaling symmetries,
\bea
\label{scaling1}
t\rightarrow at\ ,\qquad \Phi\rightarrow \Phi/a\ ,\qquad f\rightarrow f/a^2\ ,\qquad h\rightarrow a h\ ,\\
\nn\\
\label{scaling2}
r\rightarrow ar\ ,\qquad (t,x,y)\rightarrow (t,x,y)/a\ ,\qquad f\rightarrow a^2 f\ ,\qquad \Phi\rightarrow a\Phi\ .
\eea
The scaling (\ref{scaling1}) is used to set $h(r_h)=1$ and the scaling (\ref{scaling2}) is used to fix the value of the horizon coordinate, we take $r_h=1$. Then, we have three independent parameters, $\{\Phi(r_h),\eta(r_h),\theta(r_h)\}$, and only two asymptotic conditions to match. It follows that the solution is specified only by one parameter. Once the charge of the black hole has been fixed to a constant value, for example $\rho=1$, this parameter is the temperature (\ref{Hawking}). 

The superconducting black holes are dual to a field theory ground state. The thermodynamical properties of this ground state can be investigated using the bulk solutions. We briefly review how the field theory thermodynamics is recovered from the $AdS/CFT$ correspondence. 

The entropy density $\hat{s}$ of the boundary field theory is related the surface area of the horizon through the formula,
\be
\hat{s}=\frac{2\pi r_h^2}{\kappa^2}\ .
\ee  
The other thermodynamical quantities of interest are: the energy density $\hat{\epsilon}$, the free energy $\hat{f}$, the charge density $\hat{\rho}$ and the chemical potential $\hat{\mu}$. They are related to the coefficients of the boundary expansion, (\ref{expansF}) and (\ref{expansPhi}), and are given by,
\be
\hat{\epsilon}=\frac{M}{\kappa^2}\ ,\qquad \hat{\mu}\hat{\rho}=\frac{\mu\rho}{2\kappa^2}\ ,\qquad\hat{f}=\hat{\epsilon}-T\hat{s}\ .
\ee  
In the canonical ensemble the following thermodynamical identity holds, 
\be\label{freEnergy}
\qquad\hat{f}=\hat{\epsilon}-T\hat{s}=\hat{\Omega}+\hat{\mu}\hat{\rho}\ ,
\ee 
where $\hat{\Omega}$ is the (density of) grand canonical potential. This potential is calculated from the 
action $\mathcal{S}_{ren}$ which has been introduced in Section \ref{S3}. We can directly substitute into (\ref{interm}) the values of $\OO_i^{(j)}$ explicitly given in (\ref{Coeff1}) and (\ref{Coeff2}), where $i,j=1,2$. This step yields the result,
\be
\Big(2r h(r)f(r)\Big)\Big|_{r\rightarrow\infty}\ =2r^3+\frac{1}{4}\mathcal{O}_{\eta}^2\ r -2M\ .
\ee
Therefore, in order to renormalize the action we need to add the Gibbons-Hawking term plus a boundary cosmological constant and the counterterm for a $\Delta=1$ condensate dual to $\eta(r)$,
\be
S_1=-\frac{1}{2\kappa^2} \int d^3x \sqrt{-g_{B}}\left(\ \eta\left(n^{\mu}\partial_{\mu}\eta\right)\ +\ \eta^2/2 \right)\ .
\ee
The final result gives the relation $-\mathcal{S}_{ren}=\int d^3x\ M/(2\kappa^2)$ and thus, $\hat{\Omega}=TS_{E}=-M/2$. \\
The relation (\ref{freEnergy}) can be derived from a conserved charge. The procedure to obtain the conserved charge is straightforward once the Einstein equation, $R_{xx}-\kappa^2 g_{xx}\mathcal{L}=0$, is considered. This equation is not independent from the set of equations (\ref{Eqchi})-(\ref{Eqtheta1}) and can be manipulated to prove that, 
\be
\mathcal{Q}(r)=r^2h(r)f'(r)-2rh(r)f(r)-r^2 h(r)\Phi(r)\Phi'(r)
\ee
is a constant of motion of each black hole solution \cite{Aprile:2009ai}. The quantity $\mathcal{Q}$ corresponds to the Noether charge of the scaling symmetry (\ref{scaling2}) \cite{Gubser:2009cg}. Then, the relation $\mathcal{Q}(r_h)=\mathcal{Q}(r\rightarrow\infty)$ is equivalent to the thermodynamical identity  (\ref{freEnergy}).
\section{The Superpotential}
We derive the solution (\ref{analiticT0}) from a superpotential equation. This is general procedure that has a natural derivation in supersymmetric theories. In the following we review the steps of this construction. We proceed as in \cite{Faulkner:2010fh}, setting $\Phi(r)=0$ and $f(r)=r^2$. Under certain assumptions, the main idea is simply to arrange the second order equations of motion in a set of first order equations. One of the equations introduces a new quantity $\mathcal{W}$ called superpotential. The important point is that the superpotential is defined by an equation which involves only the potential $\mathcal{P}$ and its derivatives, $\partial_{\E}\mathcal{P}$ and $\partial_{\theta}\mathcal{P}$.

We are interested in the equations (\ref{Eqg}), (\ref{Eqpsi}) and (\ref{Eqtheta1}). Defining the field variable $g(r)=f(r) h(r)^2=r^2 h(r)^2$ the equations take the following form,
\begin{align}
	\label{EnergyConstraint}
-\frac{1}{4}\E'^2-\frac{1}{4}\sinh^2\left(\frac{\E}{2}\right)\theta'^2+\frac{3}{r^2}+\frac{1}{2 g}\mathcal{P}(\E,\theta)=0\ ,&\phantom{1}&\\\nn\\
\label{EnergyConstraint1}
\E''+\left(\frac{3}{r}+\frac{g'}{2g}\right)\E'-\frac{1}{4}\sinh\E\ \theta'^2-\frac{1}{g}\partial_{\psi}\mathcal{P}(\E,\theta)=0\ ,&\phantom{1}&\\\nn\\
\label{EnergyConstraint2}
\theta''+\left(\frac{3}{r}+\frac{g'}{2g}\right)\theta'+\coth\left(\frac{\E}{2}\right)\E'\theta'-
\frac{1}{\sinh^2\left(\E/2\right)}\frac{1}{g}\partial_{\theta}\mathcal{P}(\E,\theta)=0\ .&\phantom{1}&
\end{align}\\
The first one is a constraint and can be solved for $g(r)$, 
\be
g(r)= \mathcal{P}(\E,\theta) \left(\frac{1}{2}\E'^2+\frac{1}{2}\sinh^2\left(\frac{\E}{2}\right)\theta'^2-\frac{6}{r^2}\right)^{-1} \ .
\ee
Then, substituting the above expression in the remaining equations, (\ref{EnergyConstraint1}) and (\ref{EnergyConstraint1}), we obtain two equations for the two field variables, $\E(r)$ and $\theta(r)$. The final result is, 
\begin{flalign}
\E''+\left(\frac{4}{r}-\frac{r}{4}\sinh^2\left(\frac{\E}{2}\right)\theta'^2\right)\E'-\frac{r}{4}\E'^3-\frac{1}{4}\sinh\E\ \theta'^2+
\frac{\partial_{\eta}\mathcal{P}}{\mathcal{P}}\mathcal{Z}&=0\ ,\\
\nn\\
\theta''+\left(\frac{4}{r}-\frac{r}{4}\E'^2+\coth^2\left(\frac{\E}{2}\right)\E'\right)\theta'-\frac{r}{4}\sinh^2\left(\frac{\E}{2}\right)\theta'^3+
\frac{1}{\sinh^2\left(\E/2\right)}\frac{\partial_{\theta}\mathcal{P}}{\mathcal{P}}\mathcal{Z}&=0\ ,
\end{flalign}
where,
\be
\mathcal{Z}=\left(\frac{6}{r^2}-\frac{1}{2}\E'^2-\frac{1}{2}\sinh^2\left(\frac{\E}{2}\right)\theta'^2\right)\ .
\ee
At this point, the second order equations of motion can be reduced to first order equations. This procedure is general and can be written in a formal way: given the superpotential $\mathcal{W}$ the set of three non linear first order equations is\footnote{To match the notation of \cite{Faulkner:2010fh} we look at their Lagrangian ($2.1$). Then $\E\rightarrow\sqrt{2}\E$, $\theta\rightarrow\sqrt{2}\theta$ with $\mathcal{P}\rightarrow 2\mathcal{P}$ and $\mathcal{W}\rightarrow\sqrt{2}\mathcal{W}$},
\begin{align}
\label{SuperP_eta}
\E'=&-\frac{4}{r}\frac{\partial_{\eta}\mathcal{W}}{\mathcal{W}}\ ,\\
\label{SuperP_theta}
\theta'=&-\frac{4}{r}\frac{1}{\sinh^2\left(\E/2\right)}\frac{\partial_{\theta}\mathcal{W}}{\mathcal{W}}\ ,\\
\nn\\
\label{SuperP}
\mathcal{P}=&\ 4(\partial_{\eta}\mathcal{W})^2+\frac{4}{\sinh^2\left(\E/2\right)}(\partial_{\theta}\mathcal{W})^2-3\mathcal{W}^2\ .
\end{align}
In our model the superpotential can be found by inspection of the equation (\ref{SuperP}) or directly from the supersymmetry transformations of the fermions \cite{Ceresole:2001wi}. It is given by,
\be
\mathcal{W}=\sqrt{2 \cosh^2\left(\frac{\E}{2}\right)+\frac{1}{2}\cos^2\theta\ \sinh^4\left(\frac{\E}{2}\right)}\ .
\ee
Then, it is immediate to check that 
\be
\eta(r)=2\ \mathrm{arcsinh}\ \frac{C_{\eta}}{r}\ , \qquad \theta(r)=\frac{\pi}{2}\ ,
\ee 
is a solution of the system (\ref{SuperP_eta}-\ref{SuperP_theta}). 
\bigskip

\providecommand{\href}[2]{#2}\begingroup\raggedright\endgroup

\end{document}